\documentclass[11pt]{article}
\usepackage{jheppub}
\usepackage{dsdshorthand}

\usepackage{amssymb}
\usepackage{amsmath}
\usepackage{amsthm}
\usepackage{mathtools}
\usepackage{braket}
\usepackage[usenames,dvipsnames]{xcolor}
\usepackage{tikz}
\usepackage{tikz-cd}
\usepackage[skip=.5em]{parskip}

\usetikzlibrary{decorations.markings}

\usetikzlibrary{shapes.geometric, arrows}
\usepackage{appendix}
\usepackage[utf8]{inputenc}
\usepackage{dsfont}
\usepackage{cancel}
\usepackage{comment}
\usepackage{graphicx}
\usepackage{graphbox}
\usepackage{lscape}
\usepackage{float}

\usepackage{hyperref}
\hypersetup{
	colorlinks = True,
   	citecolor = magenta,
   	linkcolor = cyan,
	urlcolor = cyan,
	bookmarks=true
}

\def\L{\mathbf{L}}
\def\R{\mathbb{R}}
\def\be{\begin{equation}}
\def\ee{\end{equation}}
\def\ba{\begin{align}}
\def\ea{\end{align}}

\def\half{\frac{1}{2}}

\def\D{\Delta}
\def\a{\alpha}
\def\Db{\overbar{\Delta}}

\def\d{\text{d}}
\newcommand   \tmn   {T_{\mu \nu}}
\newcommand   \gmnu   {g^{\mu \nu}}
\newcommand   \gmnd   {g_{\mu \nu}}

\newcommand   \mnd   {_{\mu \nu}}
\newcommand   \hg   {\,_2F_1}

\newcommand   \HX   {(X^0,X^4,X^1,X^2,X^3)}
\newcommand   \NX   {(X^{-1},X^4,X^0,X^1,X^2,X^3)}
\newcommand   \NLCX   {(X^{+},X^-,X^0,X^1,X^2,X^3)}
\newcommand    \scri {\mathscr{I}}
\newcommand{\scrhp}{\mathscr{H}^+}
\newcommand{\scrhm}{\mathscr{H}^-}

\newcommand{\overbar}[1]{\mkern 1.5mu\overline{\mkern-1.5mu#1\mkern-1.5mu}\mkern 1.5mu}
\newcommand\zb{\overbar{z}}

\title{De Sitter Light-Ray Operators}

\author[a,b]{Shounak De}
\emailAdd{sde25@sas.upenn.edu}

\author[c]{and Yakov Landau}
\emailAdd{ylandau@caltech.edu}

\affiliation[a]{Center for Particle Cosmology, Department of Physics and Astronomy,
University of Pennsylvania, Philadelphia, PA 19104, USA}

\affiliation[b]{Institute for Advanced Study, Einstein Drive, Princeton, NJ 08540, USA}

\affiliation[c]{Walter Burke Institute for Theoretical Physics, Caltech, Pasadena, California 91125, USA}

\abstract{In this work, we initiate the study of light-ray operators in four-dimensional de Sitter space focusing on null integrals of the stress tensor. In Minkowski space, the null integral of the stress tensor unifies several ostensibly different constructions, functioning simultaneously as the energy flux operator, the angular contribution to a conserved charge, the averaged null energy operator, and the light transform of the stress tensor. However, we show that the de Sitter analogs of these various interpretations do not necessarily coincide but rather lead to distinct, observer-dependent light-ray operators. We construct four such de Sitter analogs and analyze their matrix elements in a free, conformally coupled scalar theory, showing that they exhibit the expected symmetry and positivity properties.}

\begin{document}

\maketitle


\vspace{-10pt}

\section{Introduction}

For over a quarter century we have known that our universe is undergoing accelerated expansion \cite{SupernovaCosmologyProject:1997zqe, Riess_1998}, and that its far future may well be described by de Sitter space. 
Despite this, our knowledge of quantum field theory (QFT) in de Sitter space is in a much more primitive state than that of its other maximally symmetric counterparts, Minkowski and AdS spaces. 
There are at least two interesting complications that make QFT in de Sitter space subtle.
The first is the lack of clearly defined Lorentzian observables, since there is no precisely defined $S$-matrix.\footnote{For recent attempts to define $S$-matrix–like observables in de Sitter space, we refer the reader to \cite{Albrychiewicz:2020ruh,Melville:2023kgd,Donath:2024utn,Melville:2024ove,Kristiano:2025cod}.} The second is the presence of infrared (IR) divergences \cite{dsir0,dsir2,dsir3,dsir5,dsir6,dsir7,dsir8,dsir10,dsir11} (see \cite{overviewdsir1,overviewdsir9} for reviews), which are manifest in loop corrections but can often even appear at tree level in the form of secular growth.

There are regimes in which some of these questions are answered, partially resolved, or mitigated. 
The inflationary regime is one such example. 
It is naturally associated with a ``meta-observer'' who has access to all of the future conformal boundary $\scri^+$, and the questions of interest are formulated in the vicinity of $\scri^+$ (see the reviews \cite{Achucarro:2022qrl,Baumann:2022jpr} and references therein for such topics). 
This is not the regime we focus on in this work. 
Instead, we are interested in the perspective of a ``bulk observer" and in the observables that are meaningful for such an observer. 
The motivation for this is two-fold: first, this is a very plausible description of the far future of our own universe; second, observers and their associated horizons are ubiquitous features in curved spacetimes. 
There has been substantial recent work taking the observer and their horizons as a starting point in general spacetimes, notably \cite{Chandrasekaran:2022cip}.
From this perspective, our focus on de Sitter space can be viewed as the study of the simplest such universe. 
The advantage is that de Sitter is a maximally symmetric spacetime and shares familiar features, such as homogeneity, isotropy and even local timelike Killing fields, while nevertheless realizing a QFT that is organized in a rather different way from its Minkowski counterpart.

We can orient ourselves in such a spacetime by asking what kind of information an observer could, in principle, have access to. 
A natural class of such observables one could consider are \textit{light-ray operators}. In Minkowski space these operators formalize many of the physical measurements performed by real-world particle detectors; they can be made IR finite, and are well defined even in the absence of an $S$-Matrix, as is the case in conformal field theories (CFTs).

Light-ray operators have two different origins. The first is their role as energy flux operators in particle physics \cite{Sveshnikov:1995vi,Korchemsky:1999kt} (for a detailed historical overview see \cite{MoultOverview}), and the second is their appearance in the context of gravity as the central object in the average null energy condition (ANEC) \cite{ANEC1}. 
The ANEC operator has been central to many important results even in flat-space QFTs, for example in constraining the behavior of renormalization group (RG) flows \cite{Hartman:2016lgu,Hartman:2023ccw,Hartman:2023qdn,Hartman:2024xkw}. 
Energy flux operators, on the other hand, have long been powerful tools in collider physics as manifestly IR-safe observables, and have been used in some of the most precise determinations of the strong coupling $\alpha_{s}$ \cite{CMS:2024mlf}.
Hofman and Maldacena \cite{H&M} showed that expectation values of these idealized detectors can be written as integrals of correlation functions of local operators. 
This then not only gave rise to a resurgence in interest in energy flux operators in particle physics \cite{MoultOverview}, but also expanded it to a much wider field. Their construction in terms of local correlation functions allows these operators to be studied in CFTs, which do not admit an $S$-matrix. 
CFTs have been shown to possess a particularly rich space of light-ray operators: any local irrep of the conformal group integrated along a null ray is another such irrep, known as its \textit{light transform} \cite{Kravchuk_2018}. 
Moreover, in the operator product expansion (OPE) of any two light transformed operators there exists an even larger class of light-ray operators, which themselves cannot be the light transform of a local operator, but rather live on Regge trajectories associated with the light transform of a local operator \cite{Kologlu:2019mfz}. 
In interacting theories, these Regge trajectories get renormalized to account for IR divergences, giving rise to a family of theory-dependent, IR-safe detector operators \cite{interacting_light_rays}.

In summary, to address theories in de Sitter space that do not admit an $S$-matrix, we will focus on a class of observables that do not require an $S$-matrix, and yet model real-world measurements. 
In Minkowski space there exists a systematic framework for rendering such operators IR safe, but we will not attempt to implement the corresponding IR analysis in de Sitter here, leaving this to future work.

To address the question of systematically constructing appropriate light-ray operators in de Sitter space, we begin in section \ref{sec:dScoordinates} by introducing the relevant coordinate systems. 
We describe four-dimensional de Sitter space first as an embedding into the hyperbolic embedding space $\mathbb{R}^{1,4}$ in section \ref{sec:hyperbolicembeddingspace}, then into the null embedding space $\mathbb{R}^{2,4}$ in section \ref{sec:nullembeddingspacedS}, before realizing its conformal completion on the Lorentzian cylinder in section \ref{lorentzian_cylinder}.
In section \ref{sec:mink_light-ray} we review the simplest light-ray operator in Minkowski space, namely the null integral of the stress tensor (defined on the future null infinity of Minkowski space $\scri^+_{\text{M}}$), and emphasize its different \textit{equivalent} incarnations as an energy flux operator, an angular contribution to the Hamiltonian, the ANEC operator, and the light transform of the stress tensor.
In section \ref{sec:classification_DSLR}, we systematically classify the de Sitter analogs of these light-ray operators, which leads to a richer set of \textit{inequivalent}, observer-dependent constructions. 
In section \ref{sec:conformal_DSLRs} we explicitly compute matrix elements of these four distinct de Sitter light-ray operators in the free conformal theory, and show how these manifest the expected symmetry and positivity properties outlined in section \ref{sec:classification_DSLR}.
Finally, in section \ref{sec:outlook} we summarize our results and conclude with a discussion of open questions and future directions suggested by this framework.


\section{de Sitter space: coordinate systems and tilings}
\label{sec:dScoordinates}

In this section, we primarily set up conventions and definitions for the various coordinate systems that will be used to define $d$-dimensional de Sitter (dS$_d$) space. Our focus will be on four dimensional de Sitter space dS$_4$, whose Ricci scalar $R$ is given by $R=12 H^2$, with $H$ being the Hubble constant. We will introduce the following three coordinate spaces that are relevant to our analysis: 
\begin{enumerate}
    \item \textit{intrinsic} coordinates on dS$_4$, denoted by lowercase Latin and Greek letters $(t,x,\tau,\theta,\dots)$,
    \item the \textit{hyperbolic embedding space} $\R^{1,4}$, denoted by uppercase Latin letters with Greek indices $(X^\mu, Z^\nu, \dots)$, and
    \item the \textit{null embedding space} $\R^{2,4}$, denoted by uppercase Latin and Greek letters with Latin indices $(X^A, Z^B, \Xi^C, \dots)$.
\end{enumerate}


\subsection{Intrinsic dS$_4$}
\label{sec:intrinsiccoord}

\begin{figure}[h]
  \centering
  \includegraphics[width=0.42\linewidth]{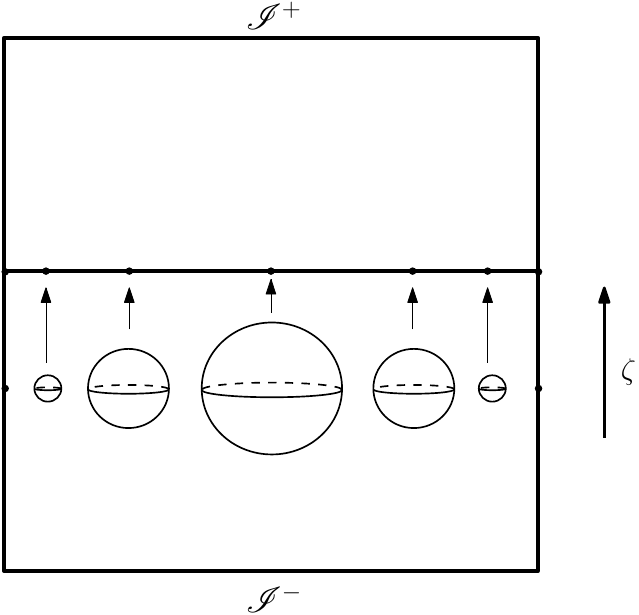}
  \caption{The usual Penrose diagram for global dS$_4$ space. Each horizontal slice represents a 3-sphere $S^3$. At $\zeta = 0$, the spatial slice is a 3-sphere $S^3$ of unit radius (in units of $H^{-1}$), which grows hyperbolically toward $\scri^\pm$, where it becomes infinitely large. Each point on such a slice represents a two-sphere $S^2$ that expands toward the center of the diagram and shrinks to a point at the left and right edges. The observer is conventionally placed on the left, at the South Pole.}
  \label{half_global}
\end{figure}

\begin{figure}[h]
  \centering
  \includegraphics[width=0.74\linewidth]{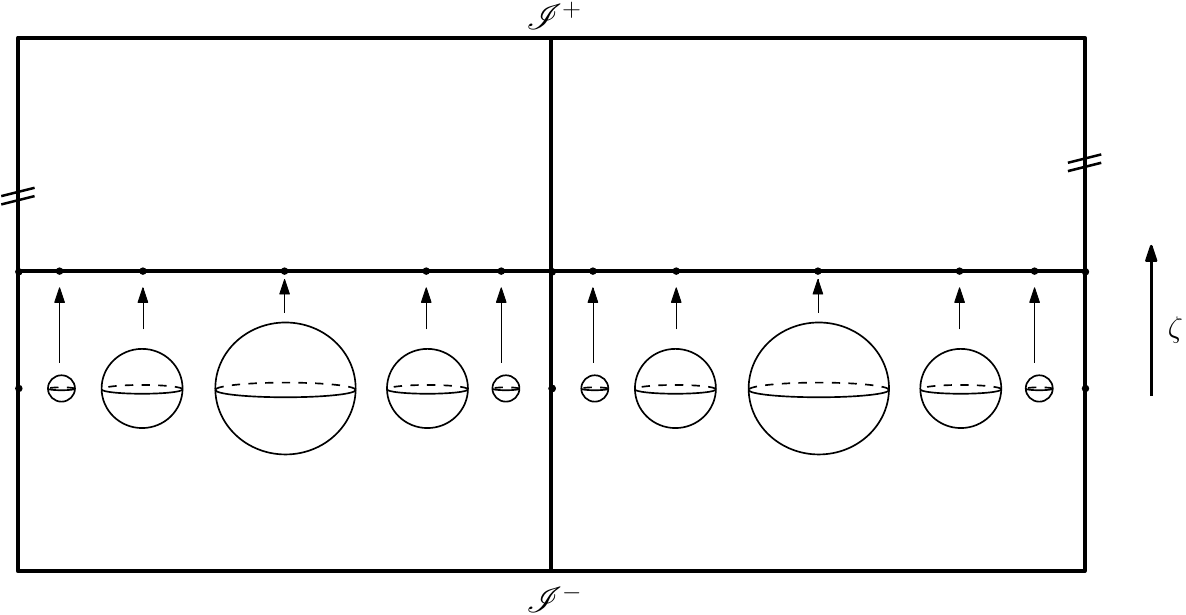}
  \caption{It will be more useful for our purposes to adopt the observer’s point of view. We therefore place the observer at the center and work with the doubled Penrose diagram, where the left and right edges are identified.}
\label{full_global}
\end{figure}

In Lorentzian signature, dS$_4$ is a maximally symmetric spacetime of constant positive curvature, with isometry group $SO(1,4)$. 
In global coordinates, it is described by the metric
\be
\label{Global_dS_metric}
\d s^2 = \ell^2(-\d \zeta^2 + \cosh^2 \zeta\: \d \Omega^2_3)~,
\ee
where $\zeta \in \R$ and $\d \Omega^2_3 = \d \theta^2 + \sin^2 \theta \ \d \Omega^2_2$ is the standard round metric on the 3-sphere $S^3$ (depicted by the Penrose diagram in \autoref{half_global}), while the radius of curvature $\ell$ is determined by the Hubble constant $\ell = H^{-1}$.
In what follows, it will be convenient for us to set $H=\ell=1$ and consequently all quantities will be measured in units of $H$ (or equivalently in units of $\ell^{-1}$).
It will be crucial for our analysis to adopt an observer’s point of view, thus breaking the homogeneity of global dS$_4$. 
To that end, we will work with the doubled Penrose diagram, whose left and right edges are identified, and with the observer’s worldline placed at the center of the diagram as shown in \autoref{full_global}.

\begin{figure}[h] 
  \centering
  \includegraphics[width=0.7\linewidth]{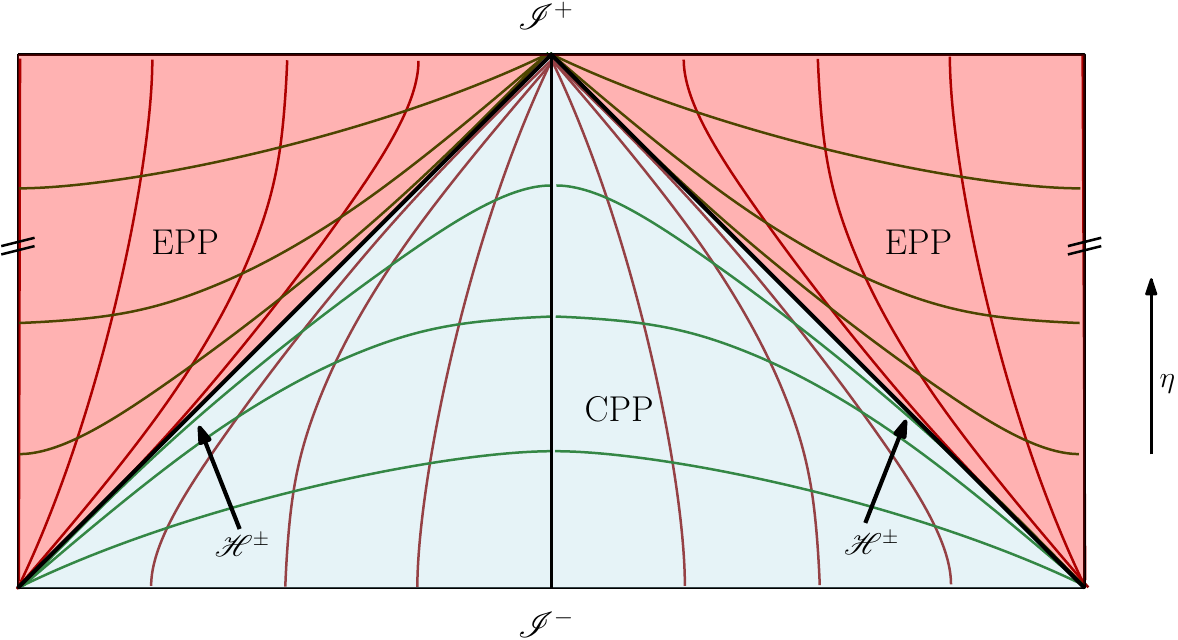}
  \caption{The Penrose diagram of dS space covered by the expanding (EPP) and contracting (CPP) Poincar\'e patches, with the observers worldline at the center of the CPP. The boundary between the patches is the $\scrhp$ of the observer at the center of the CPP and the $\scrhm$ of the observer at the center of the EPP.}
  \label{EPP_and_CPP}
\end{figure}

\paragraph{EPP/CPP Poincar\'e patches}
We will now focus on two coordinate systems that break the homogeneity of global dS$_4$ due to the presence of such an observer.
The first of these is the Poincar\'e patch(es), which cover one half of global dS$_4$ and admits spatial slices of $\R^3$. The geometry features a time-dependent scale factor that grows exponentially in the \textit{expanding} Poincar\'e patch (EPP) and shrinks exponentially in the \textit{contracting} Poincar\'e patch (CPP), corresponding to the two complementary halves of global dS$_4$. The metric in both these patches can be compactly expressed by
\be
\label{Poincare_Patch_Metric}
\d s^2 = -\d t^2 + e^{\pm 2 t}\: \d \Vec{x}^2~,
\ee
where $t \in \R$ and $\d \Vec{x}^2$ is the Euclidean metric on $\R^3$, with $\pm$ denoting the metric on the EPP/CPP respectively.
There is a useful class of coordinates on these patches known as \textit{conformal} coordinates. Setting $\mp \eta = e^{\mp t}$, the metric in \eqref{Poincare_Patch_Metric} in both the Poincar\'e patches attains the same form
\be
\label{PP_Conf_Metric}
\d s^2  = \frac{-\d \eta^2 + \d \Vec{x}^2 }{\eta^2}~.
\ee
As shown in \autoref{EPP_and_CPP}, these patches are bounded by a null horizon $\mathscr{H}^{\pm}$ and an infinitely large spatial $S^3$. 
In the case of the CPP, this $S^3$ is in the past/early-time boundary $\mathscr{I}^-$ and the null horizon $\mathscr{H}^+$ bounds its future. In the EPP, the infinitely large spatial $S^3$ is in the future/late-time boundary $\mathscr{I}^+$ and $\mathscr{H}^-$ bounds its past.

\begin{figure}[h]
  \centering
  \includegraphics[width=0.7\linewidth]{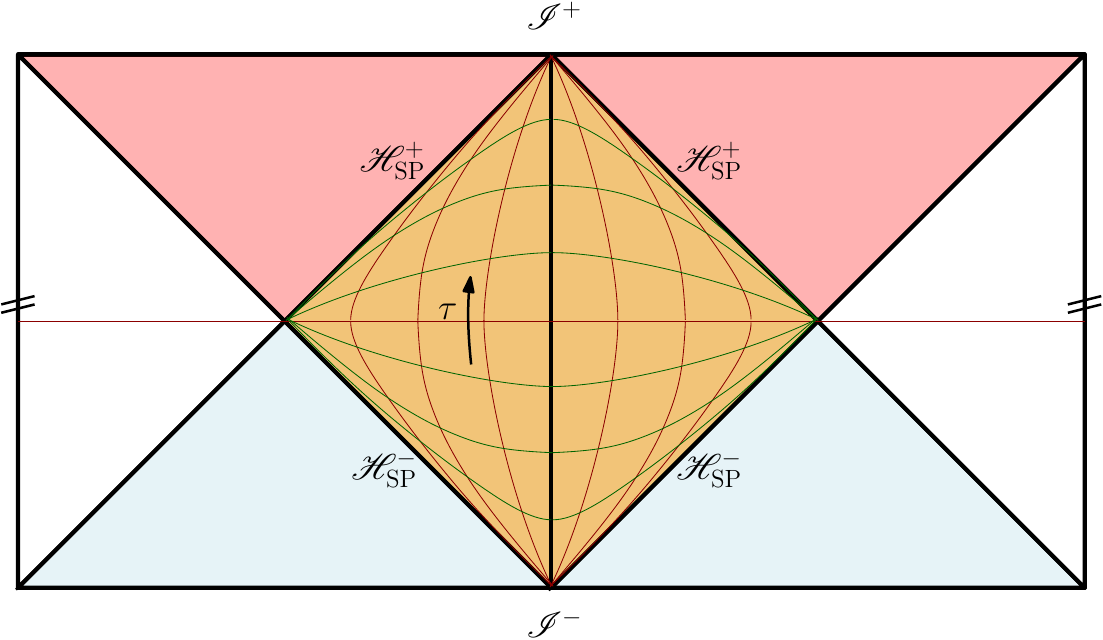}
  \caption{Static patch (gold) seen as the overlap between two Poincar\'e patches (red and blue). This is the causal patch of our observer at the center sitting at $r=0$.}
\label{static_patch}
\end{figure}

\paragraph{Static patch} We now consider \textit{static patch} coordinates, which can be seen as the intersection between the two Poincar\'e patches, and is consequently bounded by the CPP horizon $\mathscr{H}^+$ in the future and by the EPP horizon $\mathscr{H}^-$ in the past. The static patch metric is given by 
\be
\label{SP_radial_metric}
\d s^2 = - (1-r^2)\ \d \tau^2+\frac{\d r^2}{(1-r^2)}  + r^2 \d \Omega_2^2~,
\ee
where $0<r<1$ and $\d \Omega_2^2$ is the standard round metric on the $2$-sphere $S^2$. This metric is manifestly static as it is independent of the static patch time $\tau$ and is thereby endowed with a timelike Killing field $\partial_{\tau}$, which simply points in the direction of $\tau$. 
Both the past and the future cosmological/static patch horizons are at $r=1$ and will be denoted by $\mathscr{H}^-_{\text{SP}}$ and $\mathscr{H}^+_{\text{SP}}$, respectively. 
For the observer at the center (situated at $r=0$), $\mathscr{H}^{\pm}_{\text{SP}}$ is a bifurcate Killing horizon for $\partial_\tau$, and the bifurcation sphere is $S^2$ at the middle of the Penrose diagram, as shown in figure \autoref{static_patch}.

\subsection{Hyperbolic Embedding Space}
\label{sec:hyperbolicembeddingspace}

\begin{figure}[h]
  \centering
  \includegraphics[width=0.7\linewidth]{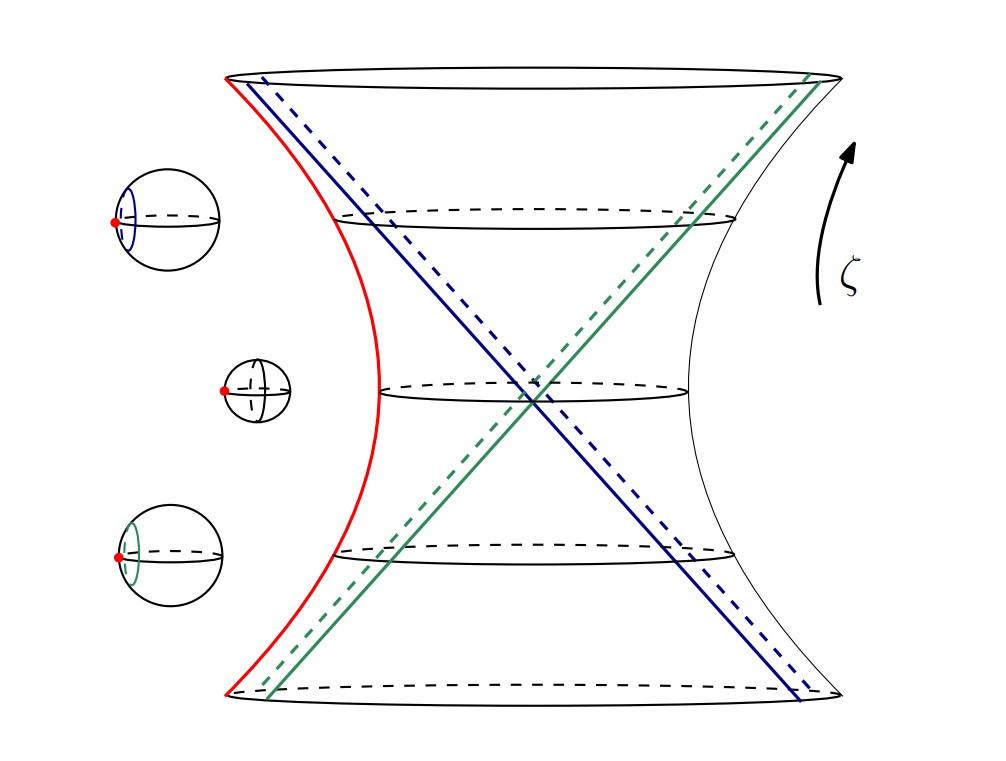}
  \caption{The embedding of dS$_4$ as a hyperboloid in $\R^{1,4}$. The observer sits at the South Pole (denoted by the red line on the left). Each horizontal cross-section represents a spatial $S^3$ slice of the global geometry. Two overlapping Poincar\'e patches define a static patch. The static patch always has the same size (of unit radius) but its portion of the overall global dS slice varies with $\zeta$. At $\zeta = 0$ it coincides with the entire left half of the $S^3$. As the global $S^3$ slices grow away from $\zeta =0$, the static patch remains a fixed-radius disk that occupies a decreasing fraction of the full sphere.}
\label{hyperbolic_embedding}
\end{figure}

\begin{figure}[h]
  \centering
  \includegraphics[width=0.7\linewidth]{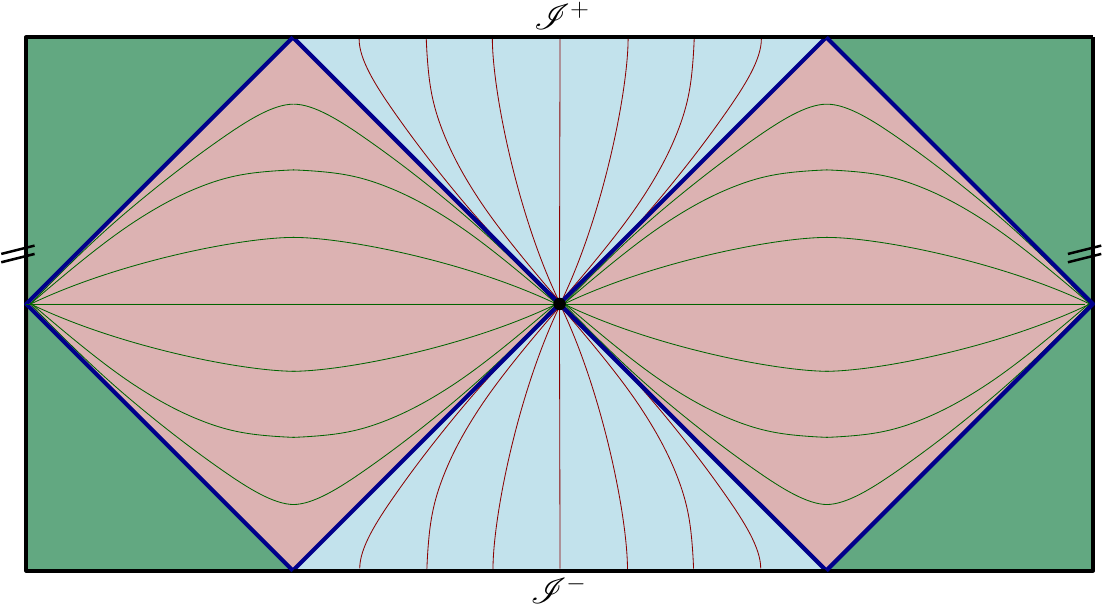}
  \caption{The observer at the center can travel along geodesics in any direction: timelike (shown in blue), spacelike (shown in red), or null in between. There also exist regions (shown in green) that cannot be reached from the origin by any geodesic.}
\label{geodesic_regions}
\end{figure}

The hyperbolic embedding space arises when dS$_4$ is viewed (\autoref{hyperbolic_embedding}) as a timelike hyperboloid living in five-dimensional Minkowski space $\R^{1,4}$ 
\be
\label{defining_hyperboloid}
\text{dS}_4 = \{ X\in \R^{1,4}:\eta\mnd X^\mu X^\nu = 1 \}~,
\ee
where $\eta\mnd = \mathrm{diag}(-1,1,1,1,1)$. 
It is worthwhile to think of this as a generalized sphere as this suggests many ways to generalize intuitions from spheres.
For example, if we consider the analytic continuation of dS$_4$ as defined in \eqref{defining_hyperboloid} to the 4-sphere $S^4$, then given any two points $X, Y \in S^{4} \subset \R^5$ their inner product 
\begin{equation}
   \sigma \equiv X\cdot Y = \cos s \in [-1,1]~,
   \label{eq:embeddingdistance}
\end{equation}
defines the so-called \textit{embedding distance},
where $s$ is the angle subtended by the great arc connecting the two points, or simply the \textit{geodesic distance}.
The embedding distance is a more natural object to work with since the $SO(5)$ isometry group of $S^4$ acts linearly on these embedding vectors and any dS invariant two-point function can be expressed in terms of the embedding distance.
In a similar fashion, we can define the $SO(1,4)$ invariant \textit{hyperbolic distance} between two points $x_i,x_j \in \text{dS}_4$ as 
\begin{equation}
    h_{ij} \equiv \eta_{\mu \nu}  X^{\mu}(x_i)  X^{\nu}(x_j)~.
    \label{eq:hyperbolicdistance}
\end{equation}
Unlike the spherical case, the exact relationship between the hyperbolic and geodesic distances will depend on whether the two points are connected by a timelike, spacelike, or null geodesic. Hence the hyperbolic distance $h$ depends on the geodesic distance $s$ as
\be
\label{hyper_dist}
h_{ij}(s) = 
\begin{cases}
\cosh s \in (1, \infty) & \text{if } s \text{ is timelike}\\
1 & \text{if } s = 0 \\
\cos s \in [-1, 1) & \text{if } s \text{ is spacelike}~.
\end{cases}
\ee 
There are also pairs of points in dS space for which $h_{ij}$ is well defined, for example with $h_{ij} < -1$, but which are not connected by any real geodesic (see \autoref{geodesic_regions}).
The regime $h_{ij} \to -\infty$ is often referred to as the cosmological limit, since it corresponds to taking both points to the late time boundary $\scri^+$ while keeping their spatial separation fixed.

We can also parameterize our tangent space $T(\text{dS}_4) \subset T(\R^{1,4})$ in a very straightforward way. 
Given any point $X^{\mu}\in \text{dS}_4$, its tangent space $T_X(\text{dS}_4)$ is defined as
\be
\label{tangent_space_definition_hyper}
T_X(\text{dS}_4) = \{Y\in T_X (\R^{1,4}): \eta\mnd X^\mu Y^\nu = 0  \} ~.
\ee
This is analogous to the case of a sphere, in that the tangent space at a point $X$ consists of all vectors in the ambient tangent space that are orthogonal to $X$ itself. 
In both cases, it is a consequence of the fact that $0 = \d (X \cdot X) = 2 X^{\mu}  \d X_{\mu}$. 
From this, it also follows that for every tensor field $H_{\mu_1 \dots \mu_n}(X)$ 
\be
X^{\mu_i} H_{\mu_1 \dots \mu_n}(X) = 0~.
\label{eq:dStensor}
\ee 
With an eye for later computations, it will also be useful to define the covariant derivative in hyperbolic embedding space as the projection of the partial derivative $\frac{\partial}{\partial X^{\nu}}$ onto dS by using the induced metric $G_{\mu \nu}$ as a projector in the manner
\be
\label{covder_def}
  \nabla_\mu H_{\nu_1 \dots \nu_l} = G_\mu^{\nu} G_{\nu_1}^{\delta_1}\cdots G_{\nu_l}^{\delta_l} \, \frac{\partial}{\partial X^{\nu}} H_{\delta_1 \dots \delta_l}~, \quad  G\mnd =  \eta\mnd -X^\mu X^\nu~.
\ee

We are now in a position to parameterize points on this hyperbolic embedding space using any of the coordinate patches described in section \ref{sec:intrinsiccoord}. For example, we can parameterize the entire hyperboloid in terms of global dS$_4$ coordinates as
\be
\label{global_hyper_embed}
\HX = (\sinh \zeta, n_i \cosh \zeta)~,
\ee
where $\vec{n}^2=1$ parametrizes the unit 3-sphere $S^3 \subset\R^4 \subset\R^{1,4}$.
The hyperbolic distance between any two points can be expressed in terms of global dS$_4$ coordinates as 
\be
\label{global_hyper_dist}
h_{ij} = -\sinh \zeta_i \sinh \zeta_j + n_i\cdot n_j \cosh \zeta_i  \cosh  \zeta_j~.
\ee 
This expression shows that when the two points differ only in their time coordinate, we have $h_{ij} = \cosh(\zeta_i-\zeta_j)$. 
If instead both points lie on the $\zeta = 0$ slice, then $h_{ij} = n_i \cdot n_j= \cos(\theta_i-\theta_j)$.

We can also parameterize portions of the hyperboloid using EPP/CPP coordinates
\be
\label{PP_hyper_embedd}
\HX = \left(\pm \sinh t + \frac{e^{\pm t}}{2}\vec{x}^2, \cosh t -\frac{e^{\pm t}}{2}\vec{x}^2,\frac{e^{\pm t}}{2}\vec{x}\right)~,
\ee
where $+$ is for the EPP, which covers the $X^0+X^4>0$ half of the hyperboloid and $-$ is for the CPP covering the other $X^0+X^4<0$ half. 
It should be emphasized that there is no unique way to choose these coordinate patches within dS space.
In the construction above, we align the coordinates with an observer located at the origin, which, to maintain correspondence with global dS space, we take to be at $\HX = (0,1,0,0,0)$. 
Accordingly, we choose the embeddings of the EPP/CPP such that their origins coincide with that of the observer (see \autoref{static_patch}). 
In this configuration, the two patches do not jointly cover the entire dS manifold but rather overlap, and their intersection corresponds to the static patch, which can be parametrized by
\be
\label{sp_hyper_embedd}
\HX = \left(\sqrt{1-r^2}\sinh \tau , \sqrt{1-r^2} \cosh \tau ,r \nhat \right)~,
\ee
where now $\nhat^2=1$ parametrizes the unit 2-sphere $\nhat \in S^2$.


\subsection{Null Embedding Space}
\label{sec:nullembeddingspacedS}

Given that dS$_4$ can be described as the hypersurface satisfying $\eta\mnd  X^\mu X^\nu = 1$ in the ambient space $\R^{1,4}$, we can equivalently view it as a projective null cone in $\R^{2,4}$. To make this manifest, let us rewrite the defining equation \eqref{defining_hyperboloid} as follows
\begin{equation}
    \eta\mnd  X^\mu X^\nu -1 =0~,
\end{equation}
which can be expressed in the homogeneous form as 
\begin{equation}
    \eta_{AB}  X^A X^B = 0~,
\end{equation}
where $X^A \in \R^{2,4}$ and one of the additional timelike coordinates is fixed to $\pm 1$ such that
\be
\label{hyper_to_null}
X^{\mu} = \HX \to X^{A} = {(X^0,X^4,\pm 1,X^1,X^2,X^3)}~.
\ee
For later convenience, we relabel the null embedding coordinates as $\NX$,\footnote{We should point that the component $X^{-1}$ should not be confused as the inverse of $X^{1}$!} and we begin with the choice $X^0=1$ (we explore the other choice in section \ref{lorentzian_cylinder}). This corresponds to a specific representative within the projective equivalence class, since the rescaling \begin{equation}
    X^A \to \l X^A~, \qquad \l \in \R^+,
\end{equation}
identifies the same point on the hyperboloid. 
With this choice in hand, all of the parameterizations introduced in section \ref{sec:hyperbolicembeddingspace} carry over directly, with the same replacement applied to each. For instance, the EPP/CPP patches in this representation can be written as
\be
\label{pp_null_embedd}
\NX = \left(\pm \sinh t + \frac{e^{\pm t}}{2}x^2, \cosh t -\frac{e^{\pm t}}{2}x^2, 1, \frac{e^{\pm t}}{2}\vec{x}\right)~,
\ee
while the static patch can be written as 
\be
\label{sp_null_embedd}
\NX = \left( \sqrt{1-r^2}\sinh \tau , \sqrt{1-r^2} \cosh \tau, 1, r \nhat \right)~.
\ee
Our notion of the hyperbolic distance \eqref{eq:hyperbolicdistance} now becomes an $SO(2,4)$ invariant distance, which we can define as the \textit{null embedding distance} $Z_{ij}$ between two points $x_i,x_j \in \text{dS}_4$ as follows
\be
\label{eq:null_distance}
Z_{ij} \equiv \eta_{AB}X^A(x_i)X^B(x_j) = \eta_{\mu \nu}X^\mu(x_i)X^\nu(x_j) - 1 = h_{ij} - 1~.
\ee
The tangent space $T_X(\text{dS}_4)$ to a point $X$ can be constructed as 
\be
\label{tangent_space_definition_null}
T_X(\text{dS}_4) = \{Y\in T_X (\R^{2,4}): \eta_{AB} X^A Y^B = 0  \} / Z-\l Z ~,
\ee
where the orthogonality is a result of $X^2 = 0 \Rightarrow X\.\d X = 0$ in hyperbolic embedding space.
The additional rescaling comes from the fact that the map $f:\R^{2,4}\to \R^{2,4}$ taking $X \to \l X$ is equivalent to the identity map. This map then induces the equality $Z \sim \l Z$ on the tangent space.

The advantage of these parameterizations, as we will soon elucidate, is that $\R^{2,4}$ provides the natural setting in which the conformal group acts linearly on dS$_4$. 
This same ambient space also furnishes the most convenient framework for describing the conformal action on four-dimensional Minkowski space. In the Minkowski case, this construction is commonly referred to as the embedding space formalism.

\subsubsection{Embedding Space: Minkowski Edition}
\label{sec:minkembeddingspace}

In this section, we briefly review the embedding space formalism \cite{Dirac:1936fq} in Minkowski space (for a more detailed discussion please refer to \cite{Kravchuk_2018, TASI_CFT_Null_Plane, Costa_2011}), which will set the stage for an analogous dS construction in the subsequent section. 
When studying a CFT in Minkowski space, there are several motivations for adopting the embedding space formalism. The first is the convenience that the conformal transformations act linearly in the embedding space. In addition, the embedding space and its tangent bundle provide a natural setting for exploring the representation theory of the conformal group. A more immediate motivation, however, is that Minkowski space itself is not closed under conformal transformations. The same is true in the Euclidean case, although there the issue is remedied by adding a single point at infinity, effectively compactifying the space into a sphere. In Lorentzian signature, by contrast, the asymptotic structure is richer as one must distinguish between spacelike, timelike, and null infinities. While spacelike ($i^0$) and timelike infinities ($i^{\pm}$) can each be represented by a single point, null infinity ($\scri^{\pm}_{\text{M}}$) requires entire surfaces. The minimal extension of Minkowski space that is closed under the conformal group is its conformal compactification, and the embedding space provides an explicit realization of this construction.

\begin{figure}[h]
  \centering
  \includegraphics[width=0.5\linewidth]{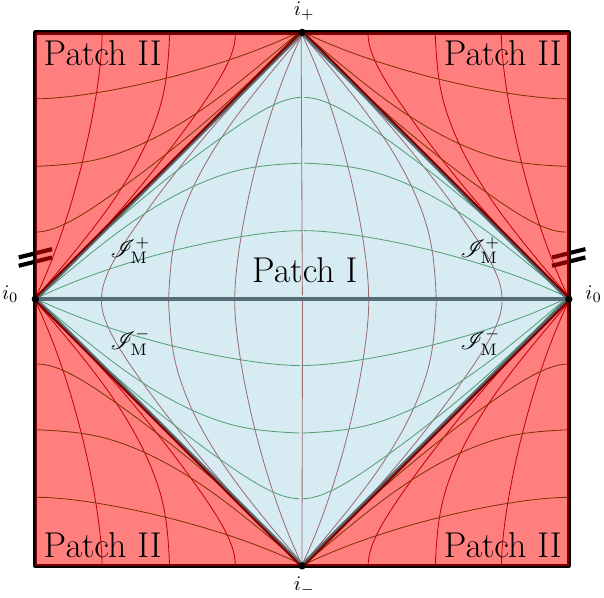}
  \caption{Two neighboring Minkowski patches on the Lorentzian cylinder shown in two dimensions, where their embedding coordinates differ by $X^+ = \pm1$. From the perspective of Patch I, the surface $X^+=0$ coincides with $\scri^+_{\text{M}}$ with the spatial infinity $X_{i^0}=(0,1,0)$ and timelike infinity $X_{i^+}=(0,-1,0)$ as its end points. The full Lorentzian cylinder is tiled by an infinite number of such Minkowski patches.}
\label{Minkowski Patches}
\end{figure}

The standard way to embed Minkowski space is to start with the projective null cone in the six-dimensional embedding space $\R^{2,4}$. 
Let us introduce lightcone coordinates $X^\pm = X^{-1} \pm X^4$ such that the metric on the embedding space $\R^{2,4}$ is
\be
X_A X^A = - X^+ X^- - (X^0)^2 + (X^1)^2 + (X^2)^2 + (X^3)^2~.
\ee
We can now identify a copy of $\R^{1,3}$ by making the gauge choice $X^+=1$, often called the Poincar\'e section.\footnote{This is not to be confused with the Poincar\'e patches of dS$_4$ introduced earlier in section \ref{sec:intrinsiccoord}.} 
Any point on the null cone with $X^+ > 0$ then admits a representative in this section. For a point $x^{\mu} \in \R^{1,3}$ , its embedding as a null vector on this Poincar\'e section defined by $X^+=1$ then has the form
\be
\label{minkowski_embedding}
\text{Patch I}: \quad x^\mu\to X^A(x) = (X^+, X^-, X^\mu) = (1,x^2,x^\mu)~,
\ee
which is depicted by Patch I in \autoref{Minkowski Patches} (in \textcolor{cyan}{blue}) on the Lorentzian cylinder. 
Moreover, the induced metric on the Poincar\'e section is 
\be\label{embed_to_mink_metric}
\d X(x)\cdot \d X(x) = -\d t^2 + \d \vec{x}^2 = \d s^2_{\R_{1,3}}~.
\ee
Furthermore, the null embedding distance between two points is expressed as
\be
\label{mink_conf_dist}
Z_{ij} \equiv X(x_i) \cdot X(x_j) = -\half (x_i - x_j)^2~.
\ee
We can also describe the tangent space as given by \eqref{tangent_space_definition_null} since Minkowski space is the same projective null cone, just parameterized differently. So given a point $x^{\mu} \in \R^{1,3}$ and a tangent vector $w^{\mu}$ at that point, we can describe its avatar $W^A(x,w)$ in the embedding space $\R^{2,4}$ as
\be
\label{flat_tangent_vectors}
W^A(x,w) = (0,2w\. x,w^{\mu})~.
\ee
We have so far neglected the points with $X^+ =0$, which correspond to the spacetime boundary $\scri_{\text{M}}^+$ of Minkowski space.
The asymptotic regions can be parameterized by using these points directly. 
To see this, one can easily verify the fact that the leading term of the embedding vector $X^A(x)$ in taking $t \to \infty$ (while keeping the spatial coordinates fixed) approaches $X_{i^+}=(0,-1,0)$, which represents future timelike infinity $i^+$. 
Similarly, spatial infinity $i^0$ is represented by $X_{i^0}=(0,1,0)$, which is the direction obtained by taking $r \to \infty$ (at fixed $t$). To identify $\scri_{\text{M}}^+$, consider any bulk point $x_0 = (t_0, r_0 \nhat)$ and a null direction $\zb =(1,\nhat)$, where $\nhat \in S^2$.
Shooting a light-ray off in the $\zb$ direction, we get
\be
x(L) = x_0 + L \zb = (t_0,r_0 \nhat) + L \zb = (t_0 + L , (r_0+L) \nhat )~,
\ee
which can be seen to approach $\scri^+_{\text{M}}$ as $L \to \infty$\footnote{One can verify that this approaches past null infinity $\scri^-_{\text{M}}$ if we instead take the limit $L \to -\infty$.}
\be 
\lim_{L \to \infty} X(x(L))\to L(0,2 \zb \cdot x_0 ,\zb^\mu) +\cO(1) = L(0,2(r_0-t_0),1,\nhat) +\cO(1)~ .
\ee
Extracting the leading order term above, we obtain the parameterization of points on $\scri^+_{\text{M}}$
\be
X_{\scri^+_{\text{M}}} = (0,2(r_0-t_0),1,\nhat) \equiv (0,-\a,1,\nhat)~,
\ee
where the real parameter $\a = 2(t_0-r_0)$ is twice\footnote{This factor of 2 is unavoidable, since the metric carries an overall factor of $1/2$. There is no particularly good convention for where it should appear, so we simply adopt this choice.} the retarded time.
We see that these limiting points are naturally parameterized by $\R \times S^2$, as is expected for $\scri^{+}_{\text{M}}$.

However, as we pass through $X^+ = 0$, we now reach the region with $X^+ < 0$, corresponding to points that cannot be connected to our previous Poincar\'e section by any positive rescaling. 
This defines a second copy of Minkowski space obtained by now choosing the section $X^+ = -1$, which we label as Patch II on the cylinder in \autoref{Minkowski Patches} (in \textcolor{red}{red})
\be
\label{minkowski_embeddingpatch2}
\text{Patch II}: \quad x^\mu\to X^A(x) = (X^+, X^-, X^\mu) = (-1,-x^2,-x^\mu)~.
\ee
In fact, each point in Patch I can be mapped to a distinct point in Patch II through the transformation $X \to \mathcal{T} X = -X$. 
The transformation $\mathcal{T}$ is defined such that for a Minkowski point $p$, the point $\mathcal{T}p$ is obtained by shooting light rays in all possible future directions from $p$ and finding the first point where they converge in the subsequent Minkowski patch.
Together, these two patches can be viewed as neighboring Minkowski patches on the Lorentzian cylinder, the universal cover of the conformal compactification of Minkowski space. 
In other words, they live on the spacetime on which $\widetilde{SO}(2,4)$, the universal cover of the Lorentzian conformal group $SO(2,4)$, acts naturally.
In fact, the full Lorentzian cylinder is tiled by an infinite number of such Minkowski patches. 


\subsubsection{Tiling the Lorentzian Cylinder with dS}
\label{lorentzian_cylinder}

One of the primary ways in which we will use the null embedding space is to relate different patches of dS space to regions of Minkowski space, allowing us to leverage results obtained in flat space within the dS context. The most useful of these relations is the one between the EPP/CPP and Minkowski space, which we alluded to above.
Let us first start with the CPP metric \eqref{Poincare_Patch_Metric} in the null embedding space, which was shown in \eqref{pp_null_embedd} to take the form
\be
\label{CPP_null_embedd}
X^A_\text{CPP}= \NX = \left(-\sinh t + \frac{e^{ -t}}{2}x^2, \cosh t -\frac{e^{ -t}}{2}x^2,  1, \frac{e^{ -t}}{2}\vec{x}\right)~.
\ee
Making the choice $e^t = +\eta$, the metric becomes \eqref{PP_Conf_Metric}, which is simply the Minkowski metric $\R^{1,3}$ with Weyl factor $\eta^{-1}$. 
If we now examine the parameterization of the CPP in the null embedding space and introduce lightcone coordinates, as we did for Minkowski space, we find
\be
\label{CPP_null_conf}
\NLCX = \left(\frac{1}{\eta},\frac{-\eta^2+\vec{x}^2}{\eta},1, \frac{\vec{x}}{\eta} \right) = \frac{1}{\eta} X^A(x) \equiv X^A_\text{CPP}(\eta, \vec{x})~.
\ee
We see that the above is equivalent to the parameterization of a point $x^{\mu} \in \R^{1,3}$ \eqref{minkowski_embedding} up to an overall factor of the Weyl factor $\eta^{-1}$. Also, the induced metric on the Poincar\'e section is the CPP metric
\be\label{embed_to_CC_metric}
\d X_\text{CPP}(\eta, \vec{x})\cdot \d X_\text{CPP}(\eta, \vec{x}) = \frac{-\d \eta^2 + \d \vec{x}^2}{\eta^2}~,
\ee
along with the fact that the null embedding distance
\be
\label{CC_conf_dist}
Z_{ij} \equiv  X_\text{CPP}(\eta_i,\vec{x}_i) \cdot X_\text{CPP}(\eta_j,\vec{x}_j) = - \frac{-(\eta_i-\eta_j)^2+(\vec{x}_i-\vec{x}_j)^2}{2 \eta_i \eta_j}~,
\ee
is the same as the Minkowski result \eqref{mink_conf_dist} up to overall Weyl factors, as expected. 
We can also now parameterize tangent space vectors, as we did in Minkowski space. At a point $x = (\eta,\vec{x})$, taking a tangent vector $w^{\mu}$ along that point is the same as \eqref{flat_tangent_vectors}
\be
\label{ds_tangent_vectors}
W(x,w) = (0,2w\. x,w^\mu)~.
\ee
Now let us again consider shooting light-rays from a generic point $x_0 = (\eta_0,r_0 \nhat)$ in the bulk of the CPP coordinate patch directed along $\zb=(1,\nhat)$ toward the future CPP horizon $\mathscr{H}^+$. This null line is parameterized by 
\be
\label{null_line}
x(L) = (\eta_0,r_0 \nhat) + L \zb = (\eta_0 + L , (r_0+L) \nhat )~,
\ee
which in the null embedding space is described by
\be 
\label{null_line_nullembeddingspace}
X_\text{CPP}(x(L)) = \left(\frac{1}{\eta_0 +L},\frac{-(\eta_0 +L)^2+(r_0+L)^2}{\eta_0 +L},1, \frac{(r_0+L)\nhat }{\eta_0 +L} \right)~.
\ee
On taking the limit $L\to \infty$, we get 
\be
\label{horizon_points1}
\lim_{L \to \infty} X_\text{CPP}(x(L)) = (0, 2x_0 \. \zb,\zb) = (0, 2(r_0 -\eta_0),1,\nhat) \equiv (0,-\a,1,\nhat)~.
\ee
As a result, we can parameterize a point on the CPP horizon $\mathscr{H}^+$ by $\a \in \R$ and a unit vector $\nhat \in S^2$. 
Putting a minus sign in front of $\a$ ensures that taking the limit $r_0 \to \oo$ coincides with $\a \to - \oo$. This means that when we parameterize the points on the CPP horizon $\mathscr{H}^+$ as
\be
\label{horizon_points}
X_{\mathscr{H}^+}(\a, \nhat) \equiv (0,-\a,1,\nhat)~,
\ee
$\a$ parametrizes a null line labeled by $\nhat$ with $\a = -\oo$ as the beginning of the null line.
For now, notice that if we map the CPP to Minkowski as suggested by \eqref{PP_Conf_Metric}, then the CPP horizon $\mathscr{H}^+$ maps to $\scri^+_{\text{M}}$ of Minkowski space.

\begin{figure}[h]
  \centering
  \includegraphics[width=0.7\linewidth]{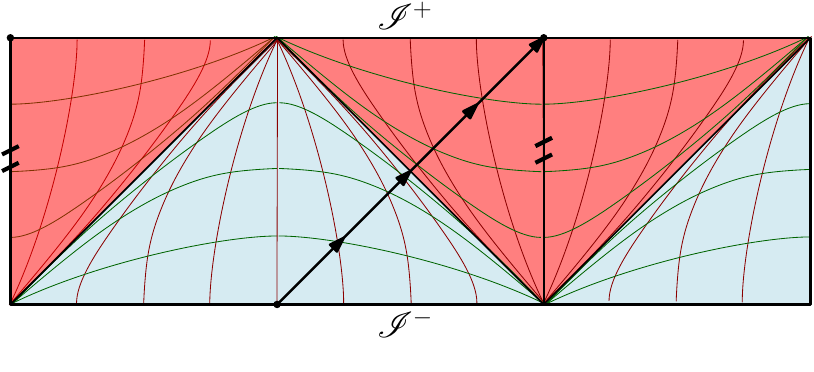}
  \caption{A null line going from the origin of one CPP (in blue), to that of a neighboring EPP (in red) on the Lorentzian cylinder. The two points in question are related by the transformation $X \to \cT X = -X$.}
\label{patch_to_patch}
\end{figure}

However, unlike $\scri^+_{\text{M}}$ of Minkowski space, as we approach the CPP horizon $\mathscr{H}^+$ ($\eta \to \infty$) we do not reach a spacetime boundary; this surface is a coordinate horizon, and one crosses smoothly into the EPP.
If our null line $x(L)$ starts from the spatial origin in the CPP at $\eta = 0$, then it must end in the EPP at its spatial origin at its $\eta = 0$, as shown in \autoref{patch_to_patch}.
To see this, consider a curve as given by \eqref{null_line} and starting at $(\eta_0,r_0) = (0,0)$
\be
\label{crossover_line}
\left(\frac{1}{L},0,1, \nhat  \right) \equiv  \left(-\b,0,1, \nhat  \right)~,
\ee
with $\b = -\frac{1}{L}$. 
Then, in the projective sense, as $\b \to -\infty$ the curve approaches $(1,0,0,\vec{0})$, which is the origin of the CPP, while as $\b = +\infty$ it approaches $(-1,0,0,\vec{0})$, which coincides with the origin of the EPP.

\begin{figure}[h]
  \centering
  \includegraphics[width=0.5\linewidth]{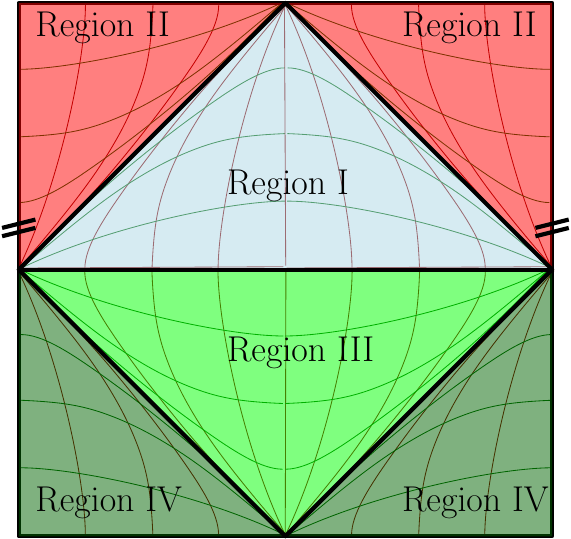}
  \caption{Two neighboring dS embeddings, and hence four associated Poincar\'e patches on the Lorentzian cylinder. Regions I and II are the CPP and EPP, respectively, of the dS patch labeled by $X^0=+1$. Regions III and IV are the EPP and CPP, respectively, of the neighboring dS patch labeled by $X^0=-1$. As in the Minkowski case, the full Lorentzian cylinder is tiled by an infinite sequence of dS embeddings, or equivalently by neighboring EPP/CPP pairs.}
\label{four_regions}
\end{figure}

We can now use this more generally to tile the Lorentzian cylinder with dS embeddings, each one built from a corresponding pair of CPP and EPP regions.
Let us begin with the CPP region already defined in \eqref{CPP_null_conf}. On the Lorentzian cylinder, this maps to the upper triangular region of Minkowski Patch I in \autoref{Minkowski Patches}, which was labeled by $X^{+}=+1$. On the Lorentzian cylinder tiled with copies of dS shown in \autoref{four_regions}, we call this Region I of the dS embedding labeled by $X^0=+1$ 
\begin{align*}
    \text{Region I}: \quad X_{\text{CPP}} &= \left( \frac{1}{\eta},\; \frac{-\eta^2 + \vec{x}^2}{\eta},\; 1,\; \frac{\vec{x}}{\eta} \right) =  \frac{1}{\eta}\left(1,\; -\eta^2 + \vec{x}^2\; \eta,\; \vec{x}\right)~~, \quad  \eta = e^t > 0~. 
\end{align*}
Next, we move to the neighboring EPP region of the same dS embedding ($X^0=+1$). We know that the conformal time $\eta$ assumes negative values $\eta = - e^{-t}$ and on the Lorentzian cylinder maps to the (upper) Minkowski Patch II of \autoref{Minkowski Patches}, which was labeled by $X^{+}=-1$. We label this EPP region as Region II of the dS embedding labeled by $X^0=+1$
\begin{align*}
    \text{Region II}:  \quad X_{\text{EPP}} &= \left( \frac{1}{\eta},\; \frac{-\eta^2 + \vec{x}^2}{\eta},\; 1,\; \frac{\vec{x}}{\eta} \right)=  \frac{1}{|\eta|}\left(-1,\; \eta^2 - \vec{x}^2,\; -\eta,\; -\vec{x}\right)~, ~  \eta = -e^{-t} < 0~,
\end{align*}
where we have factored out a positive Weyl factor since we need $X^+ = -1$. These two regions complete the tiling of the original dS embedding given by $X^0=+1$.

Equivalently, if we describe the CPP of Region I with $X^+=1$, then the neighboring EPP is in the neighboring Region II with $X^+ = -1$  obtained by $X\to \cT X = -X$. 

Let us now continue our original CPP (Region I of \autoref{four_regions}) to a new dS embedding now labeled by $X^0=-1$. In this region, $\eta=-e^{-t}$, and on the Lorentzian cylinder maps to the lower triangular region of Minkowski Patch I of \autoref{Minkowski Patches}, which was labeled by $X^{+}=+1$. This defines a new EPP region of a new dS embedding, which we call Region III on the Lorentzian cylinder
\begin{align*}
    \text{Region III}: \quad X_{\text{EPP}} &=  \left( -\frac{1}{\eta},\; \frac{\eta^2 - \vec{x}^2}{\eta},\; -1,\; \frac{\vec{x}}{\eta} \right) =  \frac{1}{|\eta|}\left(1,\; -\eta^2 + \vec{x}^2,\; \eta,\; \vec{x}\right) ~~, \quad  \eta = -e^{-t} < 0~. 
\end{align*}
We can now continue the above EPP to its neighboring CPP, which defines our final region on the cylinder. 
The conformal time in this region is positive $\eta = e^{t}$ and on the Lorentzian cylinder maps to the (lower) Minkowski Patch II of \autoref{Minkowski Patches}, which was labeled by $X^{+}=-1$. This defines the CPP region of the new dS embedding ($X^0=-1$), which we call Region IV on the Lorentzian cylinder
\begin{align*}
    \text{Region IV}: \quad X_{\text{CPP}} &=  \left( -\frac{1}{\eta},\; \frac{\eta^2 - \vec{x}^2}{\eta},\; -1,\; \frac{\vec{x}}{\eta} \right) =  \frac{1}{\eta}\left(-1,\; \eta^2 - \vec{x}^2,\; -\eta,\; -\vec{x}\right) ~~, ~~  \eta = e^t > 0~. 
\end{align*}

In summary, both Minkowski space and dS space admit conformal completions on the Lorentz-ian cylinder, described by the projective null cone in $\R^{2,4}$. 
For Minkowski space, each embedding is specified by a choice of $X^+$, while for dS space it is specified by a choice of $X^0$. 
The CPP of dS space is conformally related to the $t \geq 0$ half of Minkowski space, while the EPP is related to the $t \leq 0$ half. 
For example, the Region I CPP of the first dS embedding is conformally related to the $t \geq 0$ half of Minkowski space Patch I, while the Region III EPP of the second dS embedding is related to the $t \leq 0$ half of the same Minkowski Patch I. 
Thus, a complete embedding of dS overlaps two neighboring Minkowski embeddings, and, conversely, a complete embedding of Minkowski space overlaps two neighboring dS embeddings.  

\section{Minkowski light-ray operators: a quick primer}
\label{sec:mink_light-ray}

Light-ray operators arise naturally in many contexts in Minkowski space \cite{MoultOverview}, and many features have been generalized to spaces with Minkowski-like asymptotics \cite{gravity_light_rays}.
Each of these constructions has its own generalizations, but perhaps the simplest and most instructive of these operators is the null integral of the stress tensor.
In this section we will focus on this operator, review its definition, and briefly discuss its importance from various perspectives; much of this can be found in \cite{TASI_CFT_Null_Plane,MoultOverview,H&M}. 

\paragraph{The energy flux operator}
Let us begin with the most physical interpretation of a light-ray operator that one can realize in a flat-space QFT, the energy flux operator $\cE(\nhat)$ \cite{Sveshnikov:1995vi,Korchemsky:1999kt}. 
$\cE(\nhat)$ measures the flux of energy through a point $\nhat \in S^2$, the celestial sphere, and can be expressed as the time integral of the stress tensor in the limit of large radius as 
\begin{figure}[h]
  \centering
  \includegraphics[width=0.52\linewidth]{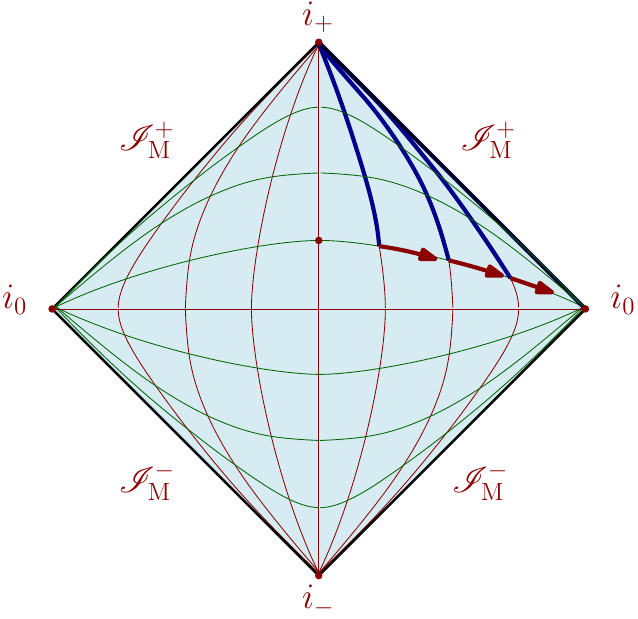}
  \caption{The energy flux operator $\cE(\nhat)$ in Minkowski space shown as a limiting procedure in two dimensions. A calorimeter (i.e. a stress-tensor insertion) placed at fixed radial distance $r$, integrated over all time, and then pushed out to $\scri^+_{\text{M}}$ as defined by the integral \eqref{E_op_def}.}
\label{energyfluxop}
\end{figure}
\be
\label{E_op_def}
\cE(\nhat)  = \lim_{r\to \oo } r^2 \int_0^\oo \d t \, T_{i \nu}(t, r \nhat) \, \nhat^i \, \xi^\nu~,
\ee
where $\xi^{\mu}=(1,0,0,0)$ is the timelike Killing vector, which becomes null at $\scri^+_{\text{M}}$.
Thus, the operator $\cE(\nhat)$ can be thought of as placing a calorimeter (i.e.~a stress tensor insertion) at some finite radial distance $r$ from the spatial origin, with the interpretation that $r^2 T^0_{~i}$ measures the local energy flux in the spatial $\nhat$ direction.
Integrating over time collects all the radiation deposited at the point $\nhat$ on the celestial sphere.
Finally, taking the limit of large radius (the factor $r^2$ ensures finiteness of the large radius limit $r \to \oo$), one pushes the calorimeter far away from the center of Minkowski space. As seen on the Penrose diagram in \autoref{energyfluxop}, this ensures that the calorimeter asymptotes to future null infinity $\scri^+_\text{M}$.

Let us now try to express the integral \eqref{E_op_def} in a more covariant language utilizing the Minkowski embedding space $\R^{2,4}$ (section \ref{sec:minkembeddingspace}).
The process of taking the large radius limit in \eqref{E_op_def} is a bit subtle, and it is convenient to introduce the null tangent vectors
\be
z^{\mu} = (1,-\nhat)~, \qquad \zb^{\mu} = (1,\nhat)~,
\ee
such that we can realize $\frac{1}{4}(z^\mu z^\nu-\zb^\mu \zb^\nu)\tmn = \nhat^i T^0_{~i}$, which is the right-hand side of \eqref{E_op_def}. 
Now considering the starting point of the integral to be $X^A(x) = (1,x^2,x^\mu)$, if we first take the $r\to \oo$ limit, we end up at $X_{i_0} = (0,1,0)$ and lose any time dependence.
Instead, we should use retarded time $\a = 2(t-r)$ and take the $r\to \oo$ limit, while keeping $\a$ fixed. 
In the embedding space, if we express our null tangent space vectors $z,\zb$ as suggested by (\ref{flat_tangent_vectors}), we can see that $Z(z)$ grows linearly in $r$ in the large $r$ limit, while $\overbar{Z}(\zb)$ does not. 
We can therefore drop the $\zb$ terms in the large $r$ limit, which means that the integral is expressed as
\be
\label{energy_flux_bulk}
\cE(\nhat)  =  \lim_{r\to \oo} \frac{r^2}{8}\int_{-2r}^{\oo} \d \a \, Z^A Z^B T_{AB} \left(X\left(\frac{\a}{2} +r,r \nhat\right)\right)~.
\ee
Equipped with this definition of energy flux operators, the natural observables to study are its correlation functions of the form
\be
\braket{\cE(\nhat_1) \cE(\nhat_2) \cdots \cE(\nhat_n)}~,
\ee
which probe the angular distribution of energy and encode, for example, the jet structure produced in a given scattering event.
We can compute this in a generic state $\ket{\Psi}$, created for example by local scalar $\phi$ operators by computing 

\be
\label{H&M_expectation}
\braket{\Psi|\cE(\nhat) |\Psi} = \int \d^4 x_1 \d^4 x_2\Psi^*(x_1)\Psi(x_2) \braket{0|\phi^*(x_1)\cE(\nhat) \phi(x_2)|0} ~.
\ee
One can naturally generalize this to other integrals of measurable quantities, for example, one can consider higher moments of the energy deposited at a point in the celestial sphere, see \cite{TASI_CFT_Null_Plane} for example. 

\paragraph{Angular contribution to the Hamiltonian} There is a complementary interpretation of $\cE(\nhat)$: it represents the angular distribution of a conserved charge.
To see this, note that integrating $\cE(\nhat)$ over the celestial sphere at infinity, the $S^2 \subset \scri^+_{\text{M}}$, amounts to integrating the $T_{\mu\nu} \xi^\mu$ component of the stress tensor across a codimension-one surface at future null infinity $\scri^+_{\text{M}}$. This reproduces the total Hamiltonian 
\be
\label{eq:mink_hamiltonian}
H = \int_{S^2} \d \Omega_2 \, \cE(\nhat) = \int_{\scri^+_{\text{M}}} \d\s^\nu \ T_{\mu\nu} \xi^\mu~,
\ee
where $\d \O_2$ is the volume form on the 2-sphere, and after including the $\d \a$ measure along the null direction, $\d\s^\nu$ becomes the natural volume form on $\scri^+_{\text{M}}$.
More generally, replacing $T^\mu_{~\nu} \xi^\nu$ with an arbitrary conserved current $J^{\mu}$ defines analogous charge measuring operators, and integrating these over $S^2$ yields the corresponding conserved charge \cite{H&M}.

\paragraph{The ANEC operator}
Another important light-ray operator is the averaged null energy operator or simply, the ANEC\footnote{Of course, the ANEC operator is a somewhat unfortunate name, since it should be the ANE operator, or the operator that gives rise to the ANEC inequality. However, the ANEC operator has a good ring to it, and everyone else seems to do it, so why should we deprive ourselves.} operator.
The construction begins by choosing any complete achronal null geodesic $\g$ in an arbitrary curved spacetime and integrating the stress tensor along its null direction. 
In Minkowski space, this takes the form
\be
\label{ANEC}
\mathcal{A}_{\text{M}} = \int_\g \d \a \ \tmn z^\mu z^\nu~,
\ee
where $z^{\mu}$ is the tangent to the curve $\gamma$. 

\paragraph{The light transform}
In CFTs there is another well-studied class of light-ray operators defined by a conformally invariant integral transform called the \textit{light transform} \cite{Kravchuk_2018}. 
The light transform $\L$ maps any local primary operator of dimension and spin $(\D,J)$ to a corresponding nonlocal operator with dimension and spin $(1-J,1-\D)$ by integrating it over a complete, affine, null geodesic.
Consider a local operator $\cO$ that lives in a representation labeled by $(\D,J)$.
Then, given a null ray in the $z$ direction passing through a point $y$, we can define its light transform as
\be
\label{poinc_light_transform}
\bL[\cO](x,z) = \int^{\oo}_{-\oo}  \d\a \ z^{\mu_1} \cdots z^{\mu_J} \cO_{\mu_1 \cdots \mu_J}(y+\a z)~,
\ee
where $x$ is the point at the beginning of the null line, i.e. $x= y- \oo z$. 
In the case of the spin-2 stress tensor, the corresponding light transform can be written as
\be
\label{poinc_light_transform_T}
\bL[T](x,z) = \int^{\oo}_{-\oo} \d\a \, \ z^{\mu} z^{\nu} \, \tmn(y+\a z)~.
\ee
This is easier to interpret on the Lorentzian cylinder, where we can begin the null line in one Minkowski patch and end it in another, such that $x$ is a genuine point in the interior of Minkowski space.
Letting $X$ be the corresponding point in the embedding space and $Z$ the embedding space image of the null vector $z$, we can denote a spinning operator as $\cO(X,Z) \equiv Z^A \cdots Z^B \cO_{A\cdots B}(X)$. It was shown in \cite{Kravchuk_2018} that the light transform of $\cO$ can also be expressed compactly as
\be
\label{light_transform}
\bL[\cO](X,Z) = \int_{-\oo}^{\oo} \d\a \  \cO\left(Z -\a X, -X\right)~.
\ee
Following this, the light transform of the stress tensor is simply expressed as
\be
\bL[T](X,Z) = \int_{-\oo}^{\oo} \d\a \ T \left(Z-\a X, -X\right)~.
\ee
Given two such light transformed local operators, their OPE contains more general light-ray operators that are not themselves obtained as the light transform of any local primary \cite{Kologlu:2019mfz}.
Among these, the light transform of the stress tensor plays a distinguished role, since the stress tensor is universally present in every CFT.
In holographic CFTs this operator acquires an additional interpretation: its bulk dual corresponds to a gravitational shockwave propagating along the associated null direction \cite{Shocks_superconvergence}.

We have ostensibly considered four realizations of the simplest light-ray operator: i) the energy flux operator \eqref{energy_flux_bulk}, ii) the angular contribution to the Hamiltonian \eqref{eq:mink_hamiltonian}, iii) the ANEC operator \eqref{ANEC}, and iv) the light transform of the stress tensor \eqref{poinc_light_transform_T}. 
In Minkowski space, the first two constructions are always represented by the same operator, but they do not in general coincide with the ANEC operator $\cA_{\text{M}}$ or with the light transform of the stress tensor $\bL[T]$, since the latter are defined along any complete null geodesic, whereas the energy flux operator $\cE(\nhat)$ is not. However, when all of them are placed on $\scri^+_{\text{M}}$, the four constructions agree. 
Moreover, in a conformal theory $\scri^+_{\text{M}}$ is no longer a distinguished null surface, and in that case these four notions coincide for any complete null line.
We can see this explicitly as follows. Let us begin with the light transform of the stress tensor given by (\ref{poinc_light_transform_T}), which is clearly equivalent to the ANEC operator given by (\ref{ANEC}), since the line $y+\a z$  for null $z$, is an affine parameterized complete null geodesic. 
Now, to show that the energy flux operator $\cE(\nhat)$ is the light transform of the stress tensor $\bL[T]$ on $\scri^+_{\text{M}}$ takes only a few more steps. 
In the large $r$ limit (keeping $\a$  fixed), the leading term in $X$ goes as 
\be
X \simeq r(0,-\a,\zb) = r(Z_{i_0}- \a X_{i_0})~,
\ee
where $X_{i_0} = (0,1,0)$ and $Z_{i_0} = (0,0,\zb)$. We can write the leading term of the null tangent vector $Z$ as 
\be 
Z \simeq -4r(0,1,0) =  -4rX_{i_0}~.
\ee
Substituting these into the definition (\ref{energy_flux_bulk}) of the energy flux operator $\cE(\nhat)$ and taking the large 
$r$ limit, we obtain
\begin{align}
\cE(\nhat) &= \frac{1}{8}\lim_{r\to \oo} r^2\int_{-2r}^\oo \d\a \ T(r(Z_{i_0}- \a X_{i_0}),-4rX_{i_0}) \nonumber \\
&=  \frac{1}{8}\lim_{r\to \oo} r^2\int_{-2r}^\oo \d\a \ r^{-4}(4r)^2T(Z_{i_0}- \a X_{i_0},-X_{i_0}) \nonumber \\
&= 2\int_{-\oo}^\oo \d\a \  T(Z_{i_0}- \a X_{i_0},-X_{i_0}) = 2\bL[T](X_{i_0},Z_{i_0})~,
\end{align}
where in the first line we used the homogeneity in $X$ and $Z$. 
So we can see that (up to the overall factor of 2), the energy flux operator $\cE(\nhat)$ is the light transform of the stress tensor $\bL[T]$, with the integral starting at $X_{i_0}$ and going along $\scri^+_{\text{M}}$. 
Therefore, the ANEC operator $\cA_{\text{M}}$, the light transform of the stress tensor $\bL[T]$, and the energy flux operator $\cE(\nhat)$ are in fact the same operator when defined on $\scri^+_{\text{M}}$.

In summary, the simplest light-ray operator given by the null integral of the stress tensor along a complete null line in $\scri^+_{\text{M}}$ is simultaneously an energy flux operator, an angular contribution to the Hamiltonian, the ANEC operator, and, in conformal theories, the light transform of the stress tensor. 
Each of these perspectives suggests its own natural generalizations. 
From the conformal viewpoint one is led to consider arbitrary light transformed operators and the broader class of light-ray operators they generate, while from the detector viewpoint one may generalize to other conserved currents or to higher moments of their fluxes. In all cases, the energy flux operator provides the basic and most physical example, and its various interpretations open different doors to broader classes of similar operators.


\section{de Sitter light-ray operators}

We now turn towards the main question of our paper: what are the analogs of the energy flux operator as one passes from Minkowski to dS space? 
We will begin in section \ref{sec:setupccs} by setting up our stage, which will be free scalar field theory in dS$_4$, and then note the specific properties of the conformal mass case. After appropriately defining the stress tensor, we will classify the four inequivalent generalizations of the energy flux operator in section \ref{sec:classification_DSLR}. 
We will end in section \ref{sec:conformal_DSLRs} by explicitly computing matrix elements of the various detectors in the conformal theory.

\subsection{Setup: scalar field theory in dS$_4$}
\label{sec:setupccs}

\begin{figure}[h]
  \centering
  \includegraphics[width=0.8\linewidth]{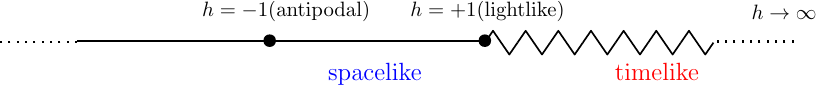}
  \caption{The three singularities of the $\hg$ function occur at $h=+1$ (lightlike separation), $h=-1$ (antipodal separation), and $h=+\infty$. The solid line indicates spacelike separated points with $h<1$, while the zigzag line indicates timelike separated points with $h>1$.}
  \label{fig:singularitiesof2f1}
\end{figure}

We consider a free scalar field theory in dS$_4$ described by the action
\be
S = -\frac{1}{2} \int~\text{d}^4x \sqrt{-g} \left(g^{\mu \nu} \nabla_{\mu} \phi \nabla_{\nu} \phi + m^2 \phi^2 + \xi R \phi^2 \right)~.
\label{eq:action}
\ee
Working in dS$_4$, we will set $R = 12$ (in terms of the unit Hubble radius) and find it convenient to define an effective mass term $\mu^2$ of the form
\be
\mu^2 \equiv m^2 + 12\xi~.
\ee
Now, the two-point Wightman function $\langle \phi(X) \f(Y) \rangle$ in the dS invariant vacuum satisfies a hypergeometric differential equation, which is best expressed in the hyperbolic embedding space $\R^{1,4}$. 
Consequently, it is a function of solely the $SO(1,4)$ invariant hyperbolic distance $h = X \cdot Y$ defined in \eqref{eq:hyperbolicdistance}.
Therefore, the two-point Wightman function is expressed simply as $\langle \phi(X) \f(Y) \rangle = G(X\. Y) \equiv G(h)$, and it is defined by the differential equation
\be
\label{2pf_hge}
(\Box - \mu^2)G(h) = (1-h^2)G''(h)-4hG'(h) -\mu^2 G(h) = 0~.
\ee
This has the solution
\be
\label{ds_hgf_2pf}
G(h) = \frac{\Gamma(\Delta)\Gamma(\overbar{\Delta})}{(4\pi)^2 \Gamma(2)}\hg \left(\Delta,\Db,2;\frac{1+h}{2}\right)~,
\ee
where the parameters $\Delta,\Db$ are defined as
\be
\label{Delta_def}
\D = \frac{3}{2} + i \nu~, \quad \quad \Db = \frac{3}{2} - i \nu~,  \quad \quad \ \nu = \sqrt{\mu^2 - \left(\frac{3}{2}\right)^2}~.
\ee
Now, note that generically the hypergeometric function $\hg(a,b,c;(1+h)/2)$ can have singularities when $h = \{-1,1,\infty\}$. 
Recalling relation \eqref{hyper_dist}, these singularities correspond to the following limits of the hyperbolic distance $h$ (depicted also in \autoref{fig:singularitiesof2f1}):
\begin{itemize}
    \item $h=1$: This is the lightcone singularity and occurs when the two points are coincident (or equivalently, the geodesic distance between them vanishes), i.e. $X\. Y = 1$.
    \item $h=-1$: This singularity corresponds to $X\. Y = -1$, which is the antipodal configuration on the sphere, meaning that the two points are maximally separated while still connected by a geodesic. 
    \item $h \to \oo$: This singularity simply corresponds to the two points being asymptotically separated in a timelike direction.
\end{itemize}

However, we choose\footnote{There is another class of solutions obtained by $h \to -h$, which arises from the $h \to -h$ symmetry of our hypergeometric differential equation \eqref{2pf_hge}. We do not take this as our solution since it does not have the lightcone singularity, but rather has its singularity at the antipode. For a more detailed discussion on this point, we refer the reader to \cite{Einhorn_2003}.} our solution $G(h)$ such that it only has singularities at $h = 1$ and $h=\oo$. The parameters\footnote{In dS$_4$, they are related to one another by the shadow transform $\Db=3-\D$, as is evident from \eqref{Delta_def}.} $\D,\Db$ are known as the scaling dimensions and they label the irreducible representations of the dS isometry group $SO(1,4)$.
There are in general three classes of irreducible representations of $SO(1,4)$: i) the \textit{principal series} corresponding to $\mu^2>(3/2)^2$ for which $\D \in \C$, ii) the \textit{complementary series} corresponding to $0<\mu^2<(3/2)^2$ for which $\D \in \R$ (but not integer valued), and iii) the \textit{exceptional/discrete series} for which $\D \in \N$ assumes non-negative integer values (for a detailed introduction please refer to \cite{Sun_representation}).

\subsubsection*{Conformally coupled scalars}
The label $\D$ is similar, but not identical, to the familiar $m^2$ in Minkowski space. 
The Wightman function exhibits different behavior for these distinct irreducible representations.
In general, $G(h)$ develops a branch cut between its singularities at $h = 1$ and $h = \oo$.
However, the analytic structure changes drastically in the special case of $\mu^2 = 2$ (or $\D=1$), which corresponds to the \textit{conformally coupled scalar}. In this case, the branch cut disappears and the Wightman two-point function is given by 
\be
\label{hyper_conf_2pf}
G(h)=\frac{\Gamma(1)\Gamma(2)}{(4\pi)^2 \Gamma(2)}\hg\left(1,2,2,\frac{1+h}{2}\right) = \frac{1}{(4 \pi)^2}\frac{2}{1-h}~.
\ee
So far in this section, our discussion has been restricted to the usage of the hyperbolic embedding space $\R^{1,4}$.
However, it will be more convenient to use the null embedding space $\R^{2,4}$.
Recall from expression \eqref{eq:null_distance} that the null embedding distance $Z$ is related to the hyperbolic distance $h$ by $h=Z+1$, with which the Wightman two-point function becomes 
\be
G(Z) = \frac{\Gamma(1)\Gamma(2)}{(4\pi)^2 \Gamma(2)}\hg\left(1,2,2,\frac{2+Z}{2}\right) = -\frac{1}{(4 \pi)^2}\frac{2}{Z}~.
\label{eq:prop_nullembedspace}
\ee
Using the expression \eqref{CC_conf_dist}, this becomes
\be
\label{null_conf_2pf}
G(Z) \equiv G(X_1,X_2) =  \frac{1}{(4 \pi)^2} \frac{4 \eta_1 \eta_2}{-(\eta_1-\eta_2)^2+(\vec{x}_1-\vec{x}_2)^2}~,
\ee
which up to the Weyl factors $\eta_1 \eta_2$, is the same as in flat space.
In fact, since the Poincar\'e patches (EPP and CPP) of dS are Weyl equivalent to Minkowski space, as established in section \ref{lorentzian_cylinder}, any correlation function of local CFT operators in dS space can be obtained from its flat-space counterpart by an appropriate Weyl rescaling via the Lorentzian cylinder.

\subsubsection*{The stress tensor}

We now move to the computation of the stress tensor for free scalar fields in dS from the action \eqref{eq:action}. 
The classical stress tensor, defined by the relation
\be
\tmn(x) = \frac{-2}{\sqrt{g}} \frac{\delta S}{\delta \gmnu}~,
\label{eq:stresstensor}
\ee
is quadratic in the field and its derivatives, and can therefore be seen as the coincident point limit of a differential operator $\Theta$ acting on the fields at separated points. 
The quantum operator will be the same with the field replaced with operators, possibly up to corrections due to regularization. 
That is to say, we can express the classical stress tensor as
\be
\label{T_as_limit}
\tmn(x) = \lim_{y \to x} \Theta\mnd (x,y) \f(x) \f(y)~,
\ee
where the differential operator $\Theta$ takes the form
\be
\label{defining_Theta}
\scalebox{0.92}{$\Theta_{\mu \nu}(x,y) = (1-2\xi) \nabla_{\mu,x} \nabla_{\nu,y} - 2\xi \nabla_{\mu,y} \nabla_{\nu,y} + g_{\mu \nu} \left[\left(2\xi - \frac{1}{2} \right) \nabla_{\rho,x} \nabla^{\rho,y} + 2 \xi \Box_x -\left(\frac{m^2}{2}+3\xi\right) \right]$}.
\ee

In the quantum theory, one can just replace the field $\f$ with the operator $\hat{\f}$ and define the stress tensor at an operational level by the limit \eqref{T_as_limit}. 
As it turns out, in the conformal case this operator is already finite. If the stress tensor is regularized using dimensional regularization, the resulting expression for the stress tensor is (the details of this derivation can be found in Appendix \ref{sec:stress_tens_reg})
\be
\label{stress_tens_with_eps}
\tmn(x)  = \lim_{\epsilon\to 0} \left( \lim_{y \to x} \Theta\mnd (x,y) \f(x) \f(y) - \frac{1}{\epsilon}\frac{\gmnd(x)}{32 \pi^2}m^2(m^2 +12 \xi -2)   \right). 
\ee
The $1/\epsilon$ term vanishes both in the massless case, $m^2 = 0$, and in the conformal case, $m^2 + 12 \xi \equiv \mu^2 = 2$. The operator is therefore finite in the case of conformally coupled scalars.


\subsection{Classification of de Sitter light-ray operators}
\label{sec:classification_DSLR}

We now turn our attention to the construction of light-ray operators in dS space. 
First, we note that there is no obvious way to generalize the construction of light-ray operators in Minkowski space (discussed in section \ref{sec:mink_light-ray}) straightforwardly to dS space.
The kinematic configuration suggested by the detector picture in Minkowski space is clearly meaningless from the perspective of global dS$_d$ space, as its spatial slices are $(d-1)$-spheres $S^{d-1}$ and therefore have empty boundaries. 
On the other hand, the conformal boundaries $\scri^\pm$ are spatial surfaces, and therefore cannot contain a null line. 
Instead, we must break global dS invariance and take the perspective of an observer. 
In other words, we take the asymptotics associated with that observer. 
The only natural null surfaces in dS are not located at the asymptotic conformal boundaries, but rather at the cosmological horizon of an observer. 
This aligns well with the idealized detector picture, since any signal or excitation placed outside an observer’s horizon is, in principle, unrecoverable.

This still does not give us a unique construction.
To see this, let us consider an observer sitting at some point in dS space, which we can naturally take to be the center of their static patch. 
They have in their vicinity a timelike Killing field, which provides them with a natural definition of energy.
If we take the conserved charge associated with this Killing field and identify it as the Hamiltonian $H$, we may extract its angular contribution exactly as we did in Minkowski space with regards to the energy flux operator $\cE(\nhat)$ in \eqref{eq:mink_hamiltonian}.
Denoting this static patch timelike Killing field
by $\xi_S = \partial_\tau$, we have
\be
\label{SP_Hamiltonian}
H = \int_{S^3} \d\s^\nu \ \tmn \, \xi_S^\mu~,
\ee
where the integral is performed over the $t = 0$ spatial slice $S^3$, and $\d\s^\nu$ is the future directed normal volume form on $S^3$. 
\begin{figure}[t]
  \centering
  \includegraphics[width=0.7\linewidth]{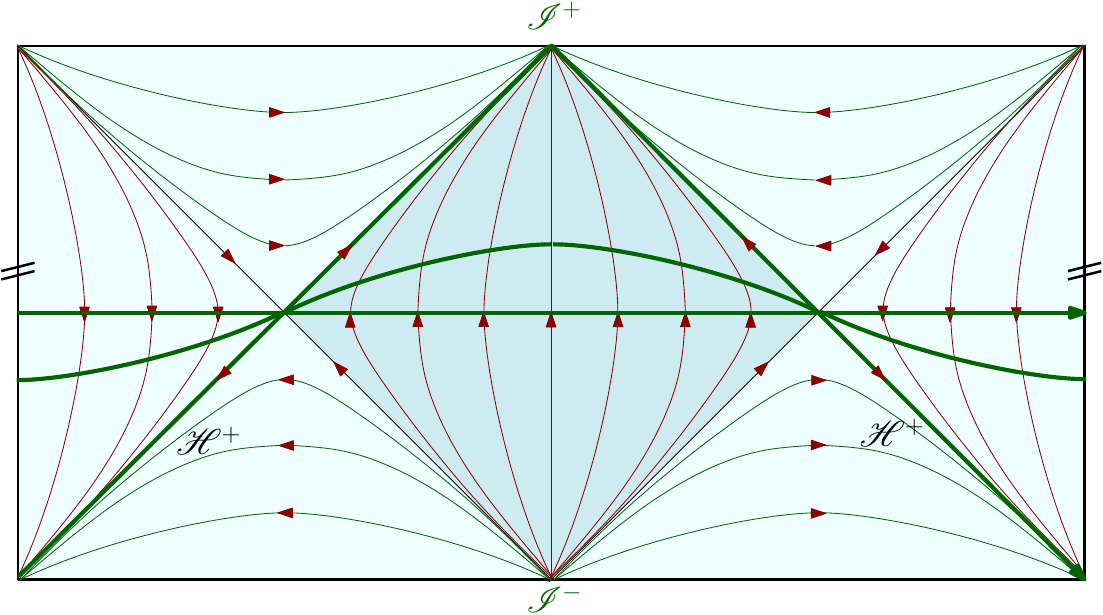}
  \caption{In this diagram, the thick \textcolor{Green}{green} segments represent not just lines, but integrals over the suppressed orthogonal $S^2$ as well. The Hamiltonian may be evaluated on any of these Cauchy slices, whether on the $t=0$ slice at the center, on the horizon $\mathscr{H}^+$, or on any interpolating surface in between. The full-horizon energy flux operator $\cE_F(\nhat)$ is defined on the entire CPP horizon $\mathscr{H}^+$.}
  \label{hamiltonian_Surfaces}
\end{figure}
We can now deform this surface up to the CPP horizon $\mathscr{H}^+$ as shown in \autoref{hamiltonian_Surfaces}. 
In that case, \eqref{SP_Hamiltonian} decomposes into an integral over the celestial sphere $S^2$ and a null direction labeled by $\a$ along which the tangent vector is $z^{\mu} = (1,-\nhat)$. 
As a result of this, we can rewrite the Hamiltonian as 
\be
\label{full_horizon_detector_definition}
H =  \int_{S^2} \d\O_2 \, \int_{-\oo}^\oo \d\a \, \tmn(\a,\nhat) \, \xi_S^\mu \, z^\nu \equiv  \int_{S^2} \d\O_2 \, \cE_F(\nhat)~.
\ee
This defines our first dS light-ray operator as an integral over the entire CPP horizon $\mathscr{H}^+$, which we call the \textit{full-horizon energy flux operator} $\cE_F(\nhat)$
\be
\label{full_horizon_detector}
\boxed{\text{Full-horizon energy flux operator}: \quad \cE_F(\nhat)  = \int_{-\oo}^\oo \d\a \, \tmn(\a,\nhat) \, \xi_S^\mu \, z^\nu}~. 
\ee
A nice feature of this construction is that under the Weyl map on the Lorentzian cylinder, the full-horizon energy flux operator \eqref{full_horizon_detector} maps to an integral along the future null infinity $\scri^+_{\text{M}}$ of Minkowski space.

\begin{figure}[h]
  \centering
  \includegraphics[width=0.7\linewidth]{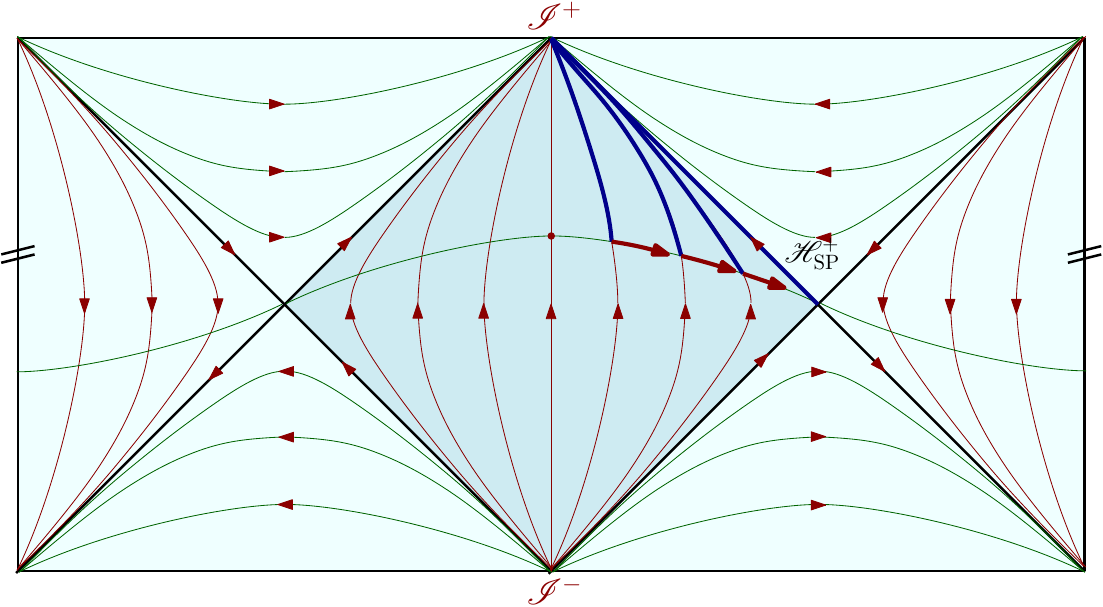}
  \caption{The observer sits at the center worldline, surrounded by their timelike Killing field. If they place a calorimeter at fixed radial distance and then move it out to the static patch horizon $\mathscr{H}^+_{\text{SP}}$, the resulting dS light-ray operator $\cE_S(\nhat)$ is integrated over $\mathscr{H}^+_{\text{SP}}$, which is half of the future CPP horizon $\mathscr{H}^+$.}
\label{Static_patch_detector}
\end{figure}

However, if we wish to follow the literal idealization of the Minkowski energy flux operator $\cE(\nhat)$ (recall \autoref{energyfluxop}), we are led to a slightly different construction in dS space. 
Analogous to the construction of $\cE(\nhat)$, consider now placing a calorimeter at fixed radial distance and allowing it to flow along a timelike geodesic with $\tau \in [0,\oo)$ denoting the static patch time associated with the conserved charge we are calling the Hamiltonian.
In the limit of large radial distance, we obtain a null line not over the full CPP horizon $\mathscr{H}^+$, but only over the portion of the horizon that intersects the static patch, as illustrated in \autoref{Static_patch_detector}.
This defines a second dS light-ray operator as an integral over the half CPP horizon or the static patch horizon $\mathscr{H}_{\text{SP}}^+$, which we call the \textit{static patch/half-horizon energy flux operator} $\cE_F(\nhat)$
\be\label{static_patch_detectors1}
\boxed{\text{Static patch/Half-horizon energy flux operator}: ~~ \cE_S(\nhat) =  \int_0^\oo  \d\a \, \tmn(\a,
\nhat) \, \xi_S^\mu \, z^\nu}~.
\ee
Note that when mapped to the Lorentzian cylinder, the above is an integral over half of $\scri^+_\text{M}$ and hence the integral \eqref{static_patch_detectors1} runs from $\a \in [0,\infty)$.

Now, neither of the above constructions, $\cE_F(\nhat)$ and $\cE_S(\nhat)$, yields the ANEC operator.
For that, we require a different operator altogether, defined by integrating the stress tensor contracted with the tangent $z^{\mu}$ to the null line taken over a complete null geodesic and not the timelike Killing field $\xi^{\mu}_S$. 
Taking this null line to lie along the CPP horizon $\mathscr{H}^+$, we can define a third dS light-ray operator called the \textit{full-horizon ANEC operator}
\be
\label{ANEC_op}
\boxed{\textrm{Full-horizon ANEC operator}: \quad \cA_F(\nhat) = \int_{-\oo}^\oo \d\a \, \tmn(\a,
\nhat) \, z^\mu \, z^\nu}~.
\ee
This operator also implements the light transform of the dS stress tensor \eqref{T_as_limit}. 
To see this, we combine two facts: first, that the ANEC operator in Minkowski space (\ref{ANEC}) is precisely the light transform of the stress tensor (\ref{poinc_light_transform_T}); and second, that the light transform, is a conformally invariant transform and hence insensitive to Weyl rescalings. Therefore, these two operators carry over identically to the conformally related setting of dS space.
Under the Weyl map from Minkowski space to the CPP (or Patch I) on the Lorentzian cylinder, the full-horizon ANEC operator $\cA_F(\nhat)$ is thus identified with the light transform of the dS stress tensor along the corresponding null generator.

Finally, there is an obvious fourth operator in the family: the analog of $\cA_F(\nhat)$, but restricted to an integral over only the future static patch horizon $\mathscr{H}^+_{\text{SP}}$, in direct analogy with \eqref{static_patch_detectors1}.
Hence we call this light-ray operator as the \textit{static patch/half-horizon ANEC-type operator} and list it below for the sake of completeness
\be
\label{SP_ANEC_op}
\boxed{\textrm{Static patch/Half-horizon ANEC-type operator}: ~\cA_S(\nhat) = \int_0^\oo \d\a \, \tmn(\a,
\nhat) \, z^\mu \, z^\nu}~.
\ee
Importantly, we note that the above detector has no natural interpretation in Minkowski space.


\subsection*{Symmetries and Full-Horizon Operators}

In the spirit of the conformal collider framework \cite{H&M}, it is natural to ask what kind of ``measurements’’ our dS light-ray operators perform and which aspects of the theory they probe.
Under the Weyl map from Minkowski space to dS conformal coordinates, the image of the timelike Killing field $\partial_t$ in Minkowski space is the conformal Killing vector (CKV) field $\partial_\eta$ associated to conformal time translation in dS space.
Said differently, conformal time translations in the CPP (and EPP) are not true isometries of dS space, but rather they are conformal isometries.
The corresponding dS light-ray operator is therefore naturally interpreted as the angular contribution to a conformally conserved charge. 
However, it is not appropriate to regard this charge as the energy, since it fails to be conserved in non-conformal theories.
In this sense, the full-horizon dS light-ray operators, $\cE_F(\nhat)$ and $\cA_F(\nhat)$, are intimately tied to the symmetry structure of dS space: $\cE_F(\nhat)$ is associated with an ordinary spacetime isometry (i.e.~static patch time translations) and can be viewed as the angular part of a conserved charge (i.e.~the Hamiltonian), while $\cA_F(\nhat)$ is associated with a conformal isometry and represents the angular part of a conformally conserved charge.

Let $Q_\xi$ and $J_\xi$ denote either of the conserved charges and its associated currents. They are related by
\be
Q_\xi = \int_\Sigma  \d \Sigma_\mu J_{\xi}^{\mu}~,
\ee
where the surface $\Sigma$ is chosen to be the CPP horizon $\mathscr{H}^+$. 
Now, the action of the Lie derivative $\cL_{\xi}$ along the associated Killing vector $\xi$ on the two-point function is given by
\be
\label{conservation_part1}
\cL_\xi(\Braket{\phi_1 \phi_2}) = i\Braket{\phi_1 [Q_\xi ,\phi_2]} = i\int_\Sigma \d \Sigma_\mu \Braket{\phi_1 [J^\mu_\xi ,\phi_2]} = i\int_\Sigma \d \Sigma_\mu \Braket{\phi_1 J^\mu_\xi \phi_2}~,
\ee
where in the final equality we have assumed that the dS vacuum is annihilated by the conserved current.
Parameterizing the CPP horizon as before using \eqref{horizon_points}, we can break up the integral on the the right-hand side of \eqref{conservation_part1} as 
\be
\label{conservation_part2}
i\int_\Sigma \d \Sigma_\mu \Braket{\phi_1 J^\mu_\xi \phi_2} = i \int_{S^2} \d\Omega_2   \Braket{\phi_1 \int_{-\oo}^{\oo} \d\a   \: z^\mu  J^\mu_\xi \phi_2} = i \int_{S^2} \d\Omega_2  \Braket{\phi_1 \cO_F(\nhat) \phi_2}~,
\ee
where $\cO_F=\cE_F, \cA_F$.
In section \ref{sec:conformal_DSLRs}, we will explicitly verify this at the level of the matrix elements of the full-horizon operators in a theory of conformally coupled scalars.

\subsection*{Computation of matrix elements: the general setup} 
Having classified the various constructions of light-ray operators in dS space, we now turn to setting up the computation of their expectation values in generic quantum mechanical states.
To begin our analysis of these operators, it is natural to start with their one-point functions. However, it is important to note that their one-point function vanishes in every theory on grounds of symmetry. $SO(1,4)$ invariance dictates that $\braket{\tmn} \sim g\mnd$, and contracting this with two null vectors necessarily gives zero.
Moreover, in order to evaluate these operators in a generic state, what we truly need are their matrix elements.
Letting $\cO$ denote any of the four operators introduced above, we want to compute 
\be
\braket{\Psi|\cO|\Psi} = \int \d^d x_1 \, \d^d x_2 \, \sqrt{g(x_1)g(x_2)} \, \Psi^*(x_1)\Psi(x_2) \braket{\phi(x_1)\cO\phi(x_2)}~.
\ee
Therefore, all we need are the matrix elements $\braket{\phi(x_1)\cO \phi(x_2)}$ where we will always fix the operator ordering via a suppressed $i\epsilon$. 
We want to express the operators $\cA_{(F,S)}(\nhat)$ and $\cE_{(F,S)}(\nhat)$ in coordinates adapted to their symmetries, namely conformal coordinates \eqref{PP_Conf_Metric} and static patch coordinates \eqref{SP_radial_metric}, respectively.
The two kinematic inputs that enter their definition are the location of the observer and the position of their horizon.
However, both of these coordinate systems become singular at the horizon. 
The way to get around this is to work with the  parameterization provided by the null embedding space $\R^{2,4}$, 
which treats the horizon in a nonsingular way.

To compute the matrix element $\braket{\phi(X_1)\cO \phi(X_2)}$, let us place the two scalar field insertions $\phi(X_1)$ and $\phi(X_2)$ at generic points $X_1, X_2$ in the interior of the CPP horizon. From \eqref{CPP_null_conf}, we know that these points are parameterized in the null embedding space by
\be
X^A_{\text{CPP}}(\eta_i, \vec{x}_i) = \frac{1}{\eta_i}\left(1,-\eta_i^2+\vec{x}_i^2 ,\eta_i ,\vec{x}_i  \right)~,
\ee
while the stress tensor sitting at a point $X_0$ on the CPP horizon $\mathscr{H}^+$ gets integrated along a null line on $\mathscr{H}^+$, which recall from \eqref{horizon_points} gets parameterized as
\be
X_{\mathscr{H}^+}(\a, \nhat) = \left(0, -\alpha, 1, \nhat \right) \equiv X_0(\a,\nhat)~.
\label{eq:pointsonthehorizon}
\ee
This implies that the tangent vector to this null line is then given by
\be
Z_{\mathscr{H}^+}(\a, \nhat) = \frac{\d X_0}{\d \a} = \left(0,-1,0,\vec{0}\right) \equiv Z_0~,
\label{eq:tangentonthehorizon}
\ee
where we note that this is simply a constant vector.
We have to now consider the two bulk Killing fields and how they are denoted at the horizon $\mathscr{H}^+$. The first is the generator $\Xi_S$ of static patch time translations $\tau \to \tau + a$, which when evaluated on the horizon is given by
\be
\Xi_S\big|_{\mathscr{H}^+} = (0, - \alpha,0,\vec{0}) = \alpha Z_0~.
\ee
The second is the CKV field $\Xi_C$ that generates conformal time translations $\eta\to \eta+ a$.
As discussed above, although this is not an isometry of dS space, it is a conformal isometry. When evaluated on the horizon, this CKV becomes
\be
\Xi_C\big|_{\mathscr{H}^+} = (0, -2, 0,\vec{0}) = 2 Z_0~.
\ee
Thus, the matrix element $\braket{\phi(X_1)\cA_F(\nhat) \phi(X_2)}$ involves contracting the stress tensor $\tmn$ with two tangent vectors $z^{\mu} z^{\nu}$, and then integrating with measure $\d\a$ according to \eqref{ANEC_op}
\be 
\braket{\phi(X_1)\cA_F (\nhat) \phi(X_2)} = 4\int_{-\oo}^{\oo} \d \a \,  \braket{\phi(X_1)T(X_0(\a, \nhat),Z_0)\phi(X_2)}~.
\label{eq:ANECmatelem}
\ee
While the matrix element $\braket{\phi(X_1)\cE_F(\nhat) \phi(X_2)}$ involves contracting the stress tensor $\tmn$ with a tangent vector $z^{\mu}$ and a Killing vector $\xi^{\mu}_S$, and then integrating with measure $\a \d\a$ according to \eqref{full_horizon_detector}
\be 
\braket{\phi(X_1)\cE_F (\nhat)\phi(X_2)} = 2\int_{-\oo}^{\oo} \d \a \,  \a \, \braket{\phi(X_1)T(X_0(\a, \nhat),Z_0)\phi(X_2)}~. 
\label{eq:energyopmatelem}
\ee

\begin{figure}[t]
  \centering
  \includegraphics[width=0.7\linewidth]{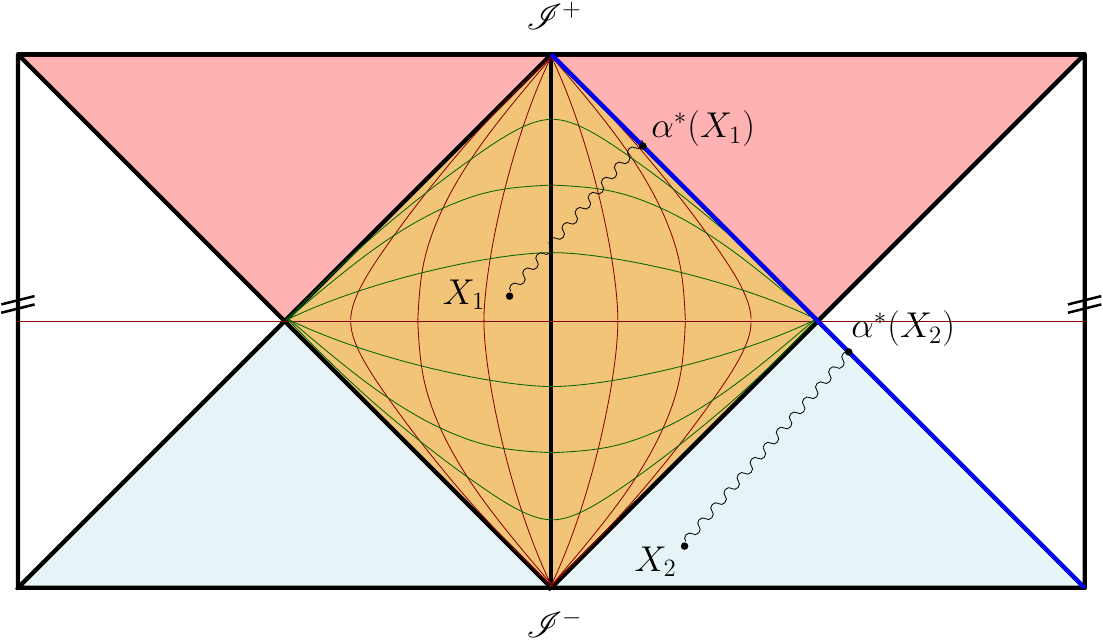}
  \caption{Two bulk points $X_1, X_2$ in dS space with their lightcones intersecting a full-horizon light-ray operator $\cO_F=\cE_F, \cA_F$ defined on the CPP horizon $\mathscr{H}^+$ (in \textcolor{blue}{blue}). Since $X_1$ lies in the static patch, we can restrict the integral \eqref{eq:genericintegrals} to begin at its singularity, so that the entire contribution is constrained to lie along the static patch horizon $\scrhp_\mathrm{SP}$.}
\label{phiTphi_singularities}
\end{figure}

Using the relations \eqref{T_as_limit} and \eqref{defining_Theta}, the common integrand in the above expressions takes the following form in the null embedding space
\begin{align}
\label{generic_phiTphi_setup}
\braket{\phi(X_1)T(X_0,Z_0) \phi(X_2)} &= \lim_{X'_0\to X_0} Z_0^A Z_0^B \Theta_{AB}\braket{\phi(X_1)\phi(X_0)\phi(X'_0)\phi(X_2)} \nonumber \\
&=2(1-2\xi)X_1\. Z_0 X_2\. Z_0 G'(Z_{01})G'(Z_{02})
\nonumber \\
&-2\xi\left((X_1\. Z_0)^2 G''(Z_{01})G(Z_{02})+(X_2\. Z_0)^2 G(Z_{01})G''(Z_{02})\right),
\end{align}
where $G(Z_{0i})\equiv G(X_0, X_i)$ is the Wightman two-point function given by \eqref{eq:prop_nullembedspace}. 
This amounts to a sum of products of hypergeometric functions, since derivatives of $\hg$ functions remain hypergeometric functions, albeit with shifted arguments.
In generic scalar QFTs in dS$_4$, we have to perform the integrals in  \eqref{generic_phiTphi_setup}, which amounts to integrating products of $\hg$ functions. 
In general, such integrals do not admit known closed-form expressions. 
We can however, make some general statements based on the analytic properties of $\hg$ functions. If we let $k = 0,1$ we can write all the integrals appearing in the matrix element $\braket{\phi(X_1)\cO \phi(X_2)}$ in the form
\be
\int \d \a \, \a^k \hg \left(a,b,c;\frac{1}{2}\left( 1+ \frac{\vec{x}_1 \. \nhat}{\eta_1}+\frac{\a}{2\eta_1} \right) \right) \hg \left(\Tilde{a} ,\Tilde{b},\Tilde{c};\frac{1}{2}\left( 1+ \frac{\vec{x}_2 \. \nhat}{\eta_2}+\frac{\a}{2\eta_2} \right) \right)~,
\label{eq:genericintegrals}
\ee
where the arguments $a,b,c$ and its tilde counterparts depend on the parameters $\Delta, \Db$.
These integrals develop a singularity when the argument $\frac{1}{2}\left( 1+ \frac{\vec{x}_i \. \nhat}{\eta_i}+\frac{\a}{2\eta_i} \right)  = 1$, which corresponds to the lightcone of $X_i$ intersecting the light-ray operator on the appropriate horizon. 
We can denote this point as $\a^*(X_i)$ as shown in \autoref{phiTphi_singularities}. 
This function has a branch cut in $\a$ that begins at $\a^*(X_i)$ and extends to $\oo$. 
For the half-horizon integrals $\cO_S=\cE_S, \cA_S$ where $\a \in [0,\infty)$, there is little additional simplification to be gained from this structure.
However, for the full-horizon integrals $\cO_F=\cE_F, \cA_F$ we can rewrite the result as an integral over the discontinuity across the branch cut, starting at either $\alpha^*(X_i)$
\be
\int_{\a^*(X_i)}^\oo \d \a \, \a^k \, \mathrm{Disc} \left[ \hg \left(a,b,c;\frac{1}{2}\left( 1+ \frac{\vec{x}_i \. \nhat}{\eta_i}+\frac{\a}{2\eta_i} \right) \right) \right]\hg \left(\Tilde{a} ,\Tilde{b},\Tilde{c};\frac{1}{2}\left( 1+ \frac{\vec{x}_j \. \nhat}{\eta_j}+\frac{\a}{2\eta_j} \right) \right). 
\ee

Interestingly, as long as at least one of the bulk points $X_1, X_2$ lies entirely within the static patch, the full-horizon operators 
becomes an integral that is entirely supported on $\scrhp_\mathrm{SP}$, as shown in \autoref{phiTphi_singularities}.
%


\subsection{Conformal de Sitter light-ray operators}
\label{sec:conformal_DSLRs}

So far, we have broadly classified the dS light-ray operators into four distinct types: the full-horizon operators $\cA_F, \cE_F$, and the static patch/half-horizon operators $\cA_S, \cE_S$.
To compute matrix elements involving these operators, we found that one must evaluate integrals of the form \eqref{generic_phiTphi_setup}, which require integrating products of hypergeometric $\hg$ functions and are, in general, quite complicated.
However, we can obtain much better analytic control by restricting our attention to the case of conformally coupled scalars, for which the two-point function was shown in \eqref{eq:prop_nullembedspace} to take the simple form
\be
G(Z_{0i}) \equiv G(X_0, X_i) = -\frac{1}{(4 \pi)^2}\frac{2}{X_0 \cdot X_i}~.
\ee
In that case, the common integrand in \eqref{generic_phiTphi_setup} simply reduces to the expression
\be
\label{phiTphi_integrand}
\braket{\phi(X_1)T(X_0,Z_0) \phi(X_2)} = -\frac{1}{96\pi^4}\frac{(X_1\. X_0 X_2\. Z_0 - X_2\. X_0 X_1\. Z_0)^2}{(X_0\. X_2)^3 (X_0\. X_1)^3}~.
\ee
One can check that this answer, up to numerical prefactors, is consistent with what is expected from conformal invariance \cite{Kravchuk_2018}. To better understand its singularity structure, let us consider the simple case where we place the two scalar operators at the two bulk points $X_1, X_2$ on the worldline of the observer. Using \eqref{CC_conf_dist}, the above expression can be written in the CPP conformal coordinates as
\be
\label{phiTphi_integrand_conf_coords_symm}
\braket{\phi(\eta_1,\vec{x}_1=0) \, T(X_0(\a, \nhat),Z_0) \, \phi(\eta_2,\vec{x}_2=0)} = -\frac{\eta_1 \eta_2}{6 \pi^4}\frac{(\eta_1 -\eta_2)^2}{(\a - 2\eta_1)^3(\a-2\eta_2)^3}~.
\ee
We see that this expression has two poles, each of third order, which appear when the integration contour crosses the lightcone of one of the bulk points. The overall factor of $\eta_1 \eta_2$ in front can be understood as the Weyl factor of the two bulk operators, while the stress tensor located on the horizon has a Weyl factor that asymptotes to unity. Said differently, we are extracting the leading term in the limit $L \gg1$, and the Weyl factor for the operator on the horizon is $\cO(L^{-1})$ (see \eqref{null_line_nullembeddingspace}). 
When the scalar operators are placed at generic bulk points $x_i=(\eta_i,\vec{x}_i)$, the numerator is again the unique conformally invariant generalization of the expression above. 
The poles then merely shift to the new locations where the integration contour intersects the new lightcones of the bulk points
\be
\label{phiTphi_integrand_conf_coords}
\braket{\phi(\eta_1,\vec{x}_1) \, T(X_0(\a, \nhat),Z_0) \, \phi(\eta_2,\vec{x}_2)} = -\frac{\eta_1 \eta_2}{6 \pi^4}\frac{z_{12}^2}{(\a +2 z_{1})^3(\a +2 z_{2})^3}~,
\ee
where we have defined
\be
z_i = \zb\. x_i~, \qquad z_{ij} = \zb_i - \zb_j~,
\ee
with $\zb=(1,\nhat)$ as before.

Having analyzed the analytic structure of the common integrand, let us now perform the $\a$-integration with the appropriate measure to compute matrix elements of the various dS light-ray operators. 
We begin with the full horizon ANEC operator $\cA_F(\nhat)$ defined in \eqref{ANEC_op}. Using \eqref{eq:ANECmatelem} along with the result for the integrand \eqref{phiTphi_integrand_conf_coords}, we obtain
\begin{align}
\label{phi_cT_phi_full}
\braket{\phi(\eta_1,\vec{x}_1) \cA_F(\nhat) \phi(\eta_2,\vec{x}_2)} &= -\frac{\eta_1 \eta_2z_{12}^2}{96 \pi^4} \int_{-\oo}^\oo \frac{\d\a}{(\a + z_{1})^3(\a + z_{2})^3} \nonumber \\
&= \frac{i}{(2\pi)^3} \frac{\eta_1\eta_2}{z_{12}^3}~.
\end{align}
Let us make two remarks on the above result.
First, we can see in this an instantiation of ANEC positivity. If we place $x_1$ and $x_2$ at the same Lorentzian point $x$, while keeping their Euclidean time separation nonzero to maintain operator ordering, we obtain
\be
\braket{\phi^*(\eta,\vec{x}) \cA_F(\nhat) \phi(\eta,\vec{x})} = \frac{i}{(2\pi)^3}\frac{\eta^2}{(-i \epsilon)^3} > 0~.
\label{eq:ANECexample}
\ee
Second, in a conformal theory, we can recognize the quantity \eqref{phi_cT_phi_full} as the integral of a conformally conserved current. 
We can show that it represents the angular contribution to the corresponding charge by integrating \eqref{phi_cT_phi_full} over a two-sphere $S^2$ and comparing the result with the Lie derivative along the Killing flow of conformal time translations, as discussed in \eqref{conservation_part1} and \eqref{conservation_part2}. 
In particular, for this CKV field $\xi_C$, we want to show that
\be
\label{killing_for_conformal}
\cL_{\xi_C}(\Braket{\phi_1 \phi_2}) = i\int_{S^2} \d \O_2 \, \Braket{\phi_1 \cA_F \phi_2}~.
\ee
The Lie derivative along the CKV field is not simply $\xi_C=\partial_{\eta_2}$. One could modify the Lie derivative to account for this explicitly, but a mathematically equivalent (and computationally simpler) approach is to evaluate both sides of \eqref{killing_for_conformal} in flat space, where the corresponding Lie derivative is just $\partial_{\eta_2}$. 
Operationally, this amounts to dropping the $\eta_1 \eta_2$ Weyl factors on both sides. In this frame, the Lie derivative reduces to a simple time derivative, and we obtain for the left-hand side
\begin{align}
\cL_{\xi_C}(\Braket{\phi_1 \phi_2}) &= -\frac{1}{4\pi^2} \partial_{\eta_2} \frac{1}{{-(\eta_1 -\eta_2)^2 +(\vec{x}_1-\vec{x}_2)^2 }} = -\frac{1}{2\pi^2}  \frac{(\eta_1 -\eta_2)}{(-(\eta_1 -\eta_2)^2 +(\vec{x}_1-\vec{x}_2)^2 )^2}~,
\end{align}
while the right-hand side of \eqref{killing_for_conformal} evaluates to
\begin{align}
i\int_{S^2} \d \O_2 \Braket{\phi_1 \cA_F \phi_2}   &= -\frac{1}{(2 \pi)^2}\int_0^\pi \frac{\sin \theta \d \theta}{(-\eta_1+\eta_2 + |\vec{x}_1 -\vec{x}_2| \cos\theta)^3} \nonumber \\
&= -\frac{1}{2\pi^2}  \frac{(\eta_1 -\eta_2)}{(-(\eta_1 -\eta_2)^2 +(\vec{x}_1-\vec{x}_2)^2 )^2}~.
\end{align}

Turning now to the case of the full-horizon energy flux operator $\cE_F(\nhat)$ defined in \eqref{full_horizon_detector}, we obtain its matrix element using the relations \eqref{eq:energyopmatelem} and \eqref{phiTphi_integrand_conf_coords} to be
\begin{align}
\braket{\phi(\eta_1,\vec{x}_1) \cE_F(\nhat) \phi(\eta_2,\vec{x}_2)} &= -\frac{\eta_1 \eta_2z_{12}^2}{96 \pi^4} \int_{-\oo}^\oo \frac{\a \d \a}{(\a + z_{1})^3(\a + z_{2})^3} \nonumber \\
&= \frac{i}{(2\pi)^3} \frac{\eta_1\eta_2 }{z_{12}^3}\frac{z_1+z_2}{2}~.
\end{align}
For this result to be related to the static patch time translation symmetry generated by $\xi_S=\partial_{\tau_2}$, we have to now show that
\be
\label{killing_for_static}
\cL_{\xi_S}(\Braket{\phi_1 \phi_2}) = i\int_{S^2} \d \O_2 \, \Braket{\phi_1 \cE_F \phi_2}~.
\ee
To do this, we begin by rewriting the conformal two-point function (\ref{null_conf_2pf}) in static patch coordinates.
We will do this in two steps. 
First, we express the conformal coordinates in a spherically symmetric form as $x_i = (\eta_i, \vec{x}_i) \to (\eta_i, \rho_i \nhat_i)$, and then change coordinates from conformal $(\eta_i,\rho_i)$ to static patch $(\tau_i,r_i)$ via $\eta_i = e^{\tau_i}(1-r_i^2)^{-1/2}$ and $\rho_i = e^{\tau_i}r_i (1-r_i^2)^{-1/2}$. 
In these coordinates, the Lie derivative of the two-point function along the timelike Killing field $\xi_S=\partial_{\tau_2}$ takes the form
\begin{align}
\cL_{\xi_S}(\Braket{\phi_1 \phi_2}) &= -\frac{1}{8\pi^2}\partial_{\tau_2}\frac{1}{-1 + \sqrt{1-r_1^2}\sqrt{1-r_2^2} \cosh(\tau_1-\tau_2)+r_1 r_2 \nhat_1\. \nhat_2} \nonumber \\
&= -\frac{1}{8\pi^2} \frac{\sqrt{1-r_1^2}\sqrt{1-r_2^2}\sinh(\tau_1-\tau_2)}{(-1 + \sqrt{1-r_1^2}\sqrt{1-r_2^2} \cosh(\tau_1-\tau_2)+r_1 r_2 \nhat_1\. \nhat_2)^2}~,
\label{static_lie}
\end{align}
which we can now compare to the right-hand side of \eqref{killing_for_static} by integrating it over $S^2$
\begin{align}
    i\int_{S^2} \d \O_2 \, \Braket{\phi_1 \cE_F \phi_2} &= -\frac{\eta_1 \eta_2}{16\pi^3}\int_{S^2} \d\O_2 \frac{z_1+z_2}{z_{12}^3}  = - \frac{\eta_1 \eta_2}{16\pi^3}\int_{S^2} \d\O_2 \frac{-\eta_1-\eta_2 + \nhat\. (\vec{x}_1 + \vec{x}_2)}{(-\eta_1+\eta_2 + \nhat\. (\vec{x}_1 - \vec{x}_2))^3} \nonumber \\
    &= \frac{i\eta_1 \eta_2}{4\pi^2}\frac{-\eta_2^2 +\vec{x}_2^2+\eta_1^2 -\vec{x}_1^2}{(-(\eta_2 -\eta_1)^2 +(\vec{x}_2 -\vec{x}_1)^2 )^2}~.
\end{align}
Rewriting this in static patch coordinates, we recover exactly the expression in \eqref{static_lie}.

The matrix element for the static patch/half-horizon ANEC-type operator $\cA_{S}(\nhat)$ defined in \eqref{SP_ANEC_op} can similarly be computed to give
\begin{align}
\braket{\phi(\eta_1,\vec{x}_1) \cA_S(\nhat) \phi(\eta_2,\vec{x}_2)} &= -\frac{\eta_1 \eta_2z_{12}^2}{96 \pi^4} \int_{0}^\oo \frac{\d\a}{(\a + z_{1})^3(\a + z_{2})^3} \nonumber \\
&=\frac{1}{192 \pi^4} \frac{\eta_1\eta_2 }{z_{12}^3} \left[ \frac{ (z_1^2- z_2^2)(z_1^2 -8 z_2 z_1+z_2^2)  }{z_1^2 z_2^2  } + 12 \log\left(\frac{z_1}{z_2}\right)\right],
\label{eq:halfhorizonanec}
\end{align}
while the corresponding result for the static patch/half-horizon energy flux operator $\cE_{S}(\nhat)$ defined in \eqref{static_patch_detectors1} gives
\begin{align}
\braket{\phi(\eta_1,\vec{x}_1) \cE_S(\nhat) \phi(\eta_2,\vec{x}_2)}  &= -\frac{\eta_1 \eta_2z_{12}^2}{96 \pi^4} \int_{0}^\oo \frac{\a \d \a}{(\a + z_{1})^3(\a + z_{2})^3} \nonumber \\
&= \frac{1}{192 \pi^4} \frac{\eta_1\eta_2 }{z_{12}^3} \left[\frac{-z_{12}(z_1^2+10 z_1 z_2 +z_2^2) }{ z_1 z_2} +6 (z_1 + z_2)\log\left(\frac{z_1}{z_2}\right) \right]~. 
\label{eq:halfhorizonenergymatelement}
\end{align}
We can see that the above matrix elements of both these half-horizon operators develop interesting new singularities at $z_i = 0$. 
The point $\a=0$ that marks the beginning of the integration contour in both the integrals \eqref{eq:halfhorizonanec} and \eqref{eq:halfhorizonenergymatelement} is denoted in null embedding space as $X_{\mathscr{H}^+}(0, \nhat) = (0,0,1,\nhat)$. 
Points in the bulk that are lightlike related to this point satisfy
\be
0 = X_i(\eta_i,\vec{x}_i) \. X_{\mathscr{H}^+}(0, \nhat) = -\eta_i + \vec{x}_i \. \nhat = z_i~.
\ee
The (logarithmic) branch cut at $z_i = 0$ thus reflects the fact that this is precisely the threshold at which the bulk points $X_i$ become lightlike separated from the integration contour of the half-horizon operator in question. 
Hence the $z_i=0$ point marks the transition between configurations where the contour does or does not intersect their lightcones.

\section{Discussions and outlook}
\label{sec:outlook}

Historically, light-ray operators have been extensively studied in Minkowski space, primarily in the context of collider physics and more recently, in CFTs and extended to generic QFTs. 
There is by now a substantial body of work analyzing these operators from many different perspectives; see, for example, the review \cite{MoultOverview} (and references therein).
In Minkowski space, the energy flux operator $\cE(\nhat)$ is in some sense canonical: broader classes of light-ray operators can be constructed from it and naturally regarded as belonging to the same family.
In this work, we have shown that an attempt to generalize these notions straightforwardly to dS space leads instead to a richer space of inequivalent constructions.
The asymptotics of dS space, together with the structure of its isometry group, allow us to classify four distinct \textit{observer-dependent} light-ray operators, as described in section \ref{sec:classification_DSLR}.
In the null embedding space $\R^{2,4}$, all of these can be written in the unified form
\begin{align}
    \cO(\nhat) = 2^{2-k} \int_{\a_i}^{\oo} \d \a \, \a^k \, T_{AB}(X_0(\a,\nhat)) Z_0^A Z_0^B~,
\end{align}
where $X_0(\a,\nhat)$ parametrizes points on the CPP horizon $\mathscr{H}^+$ (see \eqref{eq:pointsonthehorizon}), and $Z_0$ is the tangent vector along a null line on $\mathscr{H}^+$ \eqref{eq:tangentonthehorizon}, while $T_{AB}$ is the embedding space stress tensor in dS space.
The four types of dS light-ray operators then correspond to different choices of the lower limit of the integral $\a_i$ and the constant $k=0,1$, as summarized in the table below.
\begin{table}[h]
\centering
\begin{tabular}{|c|c|c|c|c|}
\hline
\textbf{Operator} $\cO(\nhat)$
& $\boldsymbol{\alpha_i}$ 
& $\boldsymbol{k}$ 
& \textbf{Horizon segment} 
& \textbf{Interpretation} \\
\hline
$\cE_F(\nhat)$ 
& $-\infty$ 
& $1$ 
& CPP horizon $\mathscr{H}^+$ 
& Full-horizon energy flux operator \\
$\cA_F(\nhat)$ 
& $-\infty$ 
& $0$ 
& CPP horizon $\mathscr{H}^+$ 
& Full-horizon ANEC operator \\
$\cE_S(\nhat)$ 
& $0$ 
& $1$ 
& Static patch horizon $\mathscr{H}^+_{\mathrm{SP}}$ 
& Static patch energy flux operator \\
$\cA_S(\nhat)$ 
& $0$ 
& $0$ 
& Static patch horizon $\mathscr{H}^+_{\mathrm{SP}}$ 
& Static patch ANEC-type operator \\
\hline
\end{tabular}

\end{table}
\vspace{-10pt}

Having classified these dS light-ray operators, we then analyzed some of their matrix elements in the simplest possible setting, namely a theory of free, conformally coupled scalars in dS$_4$, as described in section \ref{sec:conformal_DSLRs}. 
Our focus was deliberately restricted to the simplest operator in the most symmetric case, yet even in this regime the interplay between the horizon geometry, the Weyl map to Minkowski space, and the embedding space description already exhibits nontrivial structure.
At the same time, many straightforward extensions of this analysis remain unexplored, and our results should be viewed as a first step toward a broader and systematically richer family of constructions, which we discuss below:

\paragraph{General masses and interacting theories} The most immediate extensions of our analysis involve departing from the conformal point to scalars with general mass and, subsequently, to interacting theories. 
In the present work we have already seen that generic scalar QFTs in dS$_4$ give rise to detector matrix elements expressed as integrals of products of $\hg$ functions, which in general have no known closed-form expressions (see discussion below \eqref{generic_phiTphi_setup}). It would be interesting to understand to what extent special values of the mass or perturbative interactions  lead to simplifications, and whether there exist useful approximation schemes that render the space of dS light-ray correlators more tractable. 
Also, in fully interacting theories one also expects the OPE of these dS light-ray operators to become nontrivial, opening an additional direction for further study. This would then open the door to regularizing the IR divergences in a theory informed way, moving towards a dS analog of \cite{interacting_light_rays}. 

\paragraph{ANEC and positivity bounds} While we have shown an instantiation of positivity of the full horizon ANEC operator $\cA_F(\nhat)$ in \eqref{eq:ANECexample}, we leave the general proof of its positivity as future work.
In Minkowski space, light ray operators have been used to constrain the space of QFTs by constraining Wilson coefficients \cite{Shocks_superconvergence,Sharp_bounds,spinning_superconvergence}, and it would be desirable to similarly restrict the space of allowed theories in dS space. 
Our construction may provide a way to sharpen the positivity bounds of \cite{Freytsis:2022aho}, and it would be interesting to explore how these dS light-ray operators interface with recent proposals for de Sitter $S$-matrix–like observables in different regimes.

\paragraph{Cosmological collider physics} In dS space the inflationary regime has been of great interest, in which the primary observables of interest are boundary correlators defined on $\scri^{+}$ \cite{Achucarro:2022qrl,Baumann:2022jpr}. As such, it would be interesting to ask what the operators defined in our work can tell us about cosmological collider physics. Although we have defined our operators using the observer's worldline, it can equally be defined by the origin or terminus of the worldline in $\scri^\pm$, along with a choice of boost frame, and the point on the celestial sphere -- all data already defined on $\scri^\pm$. As such these operators can be entirely defined by their kinematic data on $\scri^\pm$, and can be studied in that context. In Minkowski space, the energy flux operator $\cE(\nhat)$ in the collinear limit probes jet substructure \cite{MoultOverview} and encodes detailed information about the UV theory. One might hope that suitable dS or cosmological analogs of $\cE_F(\nhat)$, or of the ANEC operator $\cA_F(\nhat)$, could probe the angular and frequency structure of late-time cosmological correlators in an analogous way. 

\paragraph{Generic curved spacetimes and quantum gravity}
Expanding on the previous direction, we note that the only kinematic input to our operators was the worldline of an observer and their associated horizons. 
These are well defined objects in generic curved spacetimes, with dS providing the simplest nontrivial example. As such, these operators capture the physics of physically realizable experiments in any curved spacetime. 
Moreover, these observables are well defined even in quantum gravity, as they are are gravitationally dressed to the observer. In particular, the half-horizon operators, associated with the static patch of the observer, sit nicely within the framework of \cite{Chandrasekaran:2022cip}, in which the generalized static patch serves as the largest regime associated to the algebra of an observer in any theory of quantum gravity.

\paragraph{Algebra of dS light-ray operators} 
Finally, it would be interesting to explore the algebraic structure of these dS light-ray operators. In Minkowski space, light-ray operators and their algebras are closely connected to asymptotic symmetries and to the structure of null infinity \cite{Cordova:2018ygx, Korchemsky:2021htm}. 
Understanding these structures could help clarify the role of dS horizons and observer dependence in any putative notion of dS holography \cite{Strominger:2001pn,Anninos:2011af,Susskind:2021omt}. 
We leave these questions, along with the many low hanging fruits outlined above, to future work.

\section*{Acknowledgments}

We would like to thank Nathan Benjamin, Cyuan-Han Chang, Hayden Lee, Kyle Lee, Ian Moult, David Simmons-Duffin, Marcus Spradlin, Anastasia Volovich, and Yixin Xu for many helpful discussions. 
In particular, we are indebted to David Simmons-Duffin for valuable guidance throughout the project and for insightful comments on the manuscript. 
SD was supported in part by the US Department
of Energy under contract DE-SC0010010 Task F, by the National Science Foundation grant NSF PHY-2309135 to the Kavli Institute for Theoretical Physics, and by funds provided by the Center for Particle Cosmology at the University of Pennsylvania. 
YL is funded in part by the National Science Foundation Graduate Research Fellowship under grant no. DGE-1745301.


\appendix
\section{Regularizing the stress tensor}
\label{sec:stress_tens_reg}

In this appendix, we will derive the regularized expression for the stress tensor \eqref{stress_tens_with_eps}. 
We will extract the divergent pieces of the stress tensor defined by the relations \eqref{T_as_limit} and \eqref{defining_Theta} using dimensional regularization. The stress tensor in this case only has one divergence, coming from its definition via point splitting. In general, interacting theories will have more divergences, but in this simple case there is only one. Additionally, even this divergence will never be an issue for us since, as we will show, the divergence is proportional to the metric; as such, when contracted with two null vectors, the divergence is zero. Another way of saying this is that the stress tensor is always a sum of two irreducible representations, the symmetric traceless part and the trace. Light-ray operators are sensitive to the traceless component while the divergence is only in the trace. Nonetheless, as we are using this operator, we include its regularization.

Let us begin by considering the insertion of the stress tensor into some generic correlation function in the following manner
\be
\Braket{\cO(z_1) \cdots \tmn(x) \cdots \cO(z_n) } = \lim_{y\to x} \Theta\mnd(x,y)\Braket{\cO(z_1) \cdots \f(y)\f(x) \cdots \cO(z_n)}~,
\ee
where we have used the defining relation \eqref{T_as_limit}.
In the free theory, for any value of the mass parameter $\mu^2$, the fact that correlation functions factorize into Wick pairs ensures that the above correlator has a particularly simple form. 
If we consider the correlation function before the $\Theta$ operator acts on it, it decomposes into a sum of Wick contractions. 
Among these terms, some will have points $x$ and $y$ each contracted with distinct points $z_i$; such contributions do not generate any short-range singularity intrinsic to the definition of the stress tensor.
The remaining terms, in which $x$ and $y$ contract with one another, will produce ultraviolet divergences.
However, in this case, the $\Theta$ operator acts only on that two-point function, and so the stress tensor factorizes out as follows
\begin{align}
\lim_{y\to x}\Theta\mnd(x,y)&\Braket{\cO(z_1) \cdots \f(y)\f(x) \cdots \cO(z_n)}  \nonumber \\
=& \lim_{y\to x} \Theta\mnd(x,y)\left[ \Braket{ \f(y)\f(x)} \right]\Braket{\cO(z_1) \cdots \cO(z_n) }  + \mathrm{finite }  \nonumber \\
=&\Braket{ \tmn(x)} \Braket{\cO(z_1) \cdots \cO(z_n) }  + \mathrm{finite}~.
\end{align}
Therefore, the divergent part of the correlator is simply the divergence of the one-point function of the stress tensor itself, which we now address. 

The calculations involving the stress tensor will be carried out in the hyperbolic embedding space following \cite{Tensor_and_Spinors}. The $\Theta$ operator given in \eqref{defining_Theta} can be defined in the hyperbolic embedding space $\R^{1,4}$ using the covariant derivative \eqref{covder_def}.
Let us consider the action of the $\Theta$ operator on a generic function $f$ of the hyperbolic distance $h \equiv X\. Y$. 
In the coincident limit $h \to 1$, we obtain
%
\be
U^A V^B\Theta _{A B}(X,Y)[ f(h)]|_{h \to 1} = -\frac{1}{2} U \cdot V ((m^2 +6\xi)f(1) +2f'(1))~,
\ee
where we have assumed that $U,V \in T_{X,Y}(\R^{1,4})$ are in the tangent space, such that $U\. X = V\. X = 0$. Consequently, we can write
\be
\langle T_{\mu\nu} \rangle_\epsilon =  -\frac{1}{2} U \cdot V ((m^2 +6\xi)G_\epsilon(1) +2G'_\epsilon(1))~,
\label{eq:divofstresstensor}
\ee
where $G_\epsilon$ is the $1/\epsilon$ divergence of the Wightman two-point function. So to regularize the stress tensor, we need to find the $1/\epsilon$ divergence of $G(h)|_{h = 1}$ and $G'(h)|_{h = 1}$. Note that identifying and subtracting the divergence in $G(h)$ is, in effect, the regularization of the $\f ^2$ operator.

\subsection{Divergence of $G(h)|_{h = 1}$}
\label{app:divofprop}

We will follow the conventions and normalizations of \cite{Chakraborty:2023qbp}. 
To extract the divergence of $G(h)$, we begin by examining the propagator $G_E$ in Euclidean\footnote{To extract the short-distance singularity, it does not matter if one is in Lorentzian or Euclidean signature.} dS$_d$, which is simply the $d$-sphere $S^d$. 
The key feature of this propagator is that it serves as the Green's function of the Klein-Gordon operator 
\begin{equation}
    G_E(x,y) = (-\Box_x + \mu^2)^{-1}~,
\end{equation}
where $\mu^2 \equiv m^2 + 12\xi$ as before. We can solve this explicitly on the $d$-sphere $S^d$ using hyperspherical harmonics 
\begin{equation}
    G_E(\sigma) = \frac{\Gamma(\alpha)}{2 \pi^{\alpha +1}}\sum_{J=0}^\infty\frac{J+\alpha}{J(J+2\alpha)+\mu^2}C^\alpha_J(\sigma)~,
\end{equation}
where $\alpha = (d-1)/2$, $\sigma$ is the Euclidean embedding distance defined in \eqref{eq:embeddingdistance}, and $C^\alpha_J$ is a Legendre function of the first kind (a Gegenbauer polynomial if $J \in \mathbb{Z}$). 
In the coincident limit $\sigma = 1$, one has the relation
\begin{equation}
    C^\alpha_J(1) = \frac{\Gamma(J+2\alpha)}{\Gamma(J+1)\Gamma(2\alpha)}~,
\end{equation}
which gives
\begin{equation}
    G_E(\sigma)|_{\sigma=1} = \frac{\Gamma(\alpha)}{2 \pi^{\alpha +1}\Gamma(2\alpha)}\sum_{J=0}^\infty\frac{J+\alpha}{J(J+2\alpha)+\mu^2}\frac{\Gamma(J+2\alpha)}{\Gamma(J+1)}~.
\end{equation}
We can simplify the expressions further by writing the denominator factor as
\begin{align}
    \frac{1}{J(J+2\alpha)+\mu^2} = \frac{1}{(J-h_+)(J-h_-)} &= \frac{\partial}{\partial \mu^2}[\mathrm{log}(J-h_+)+\mathrm{log}(J-h_-)]  \nonumber \\
    &=  \frac{\partial}{\partial \mu^2} \frac{\partial}{\partial s}[(J-h_-)^s + (J-h_+)^s]\bigg|_{s=0}~.
\end{align}
This gives us the following expression for the propagator in the coincident limit
\begin{equation}
    G_E(\sigma)|_{\sigma=1} = \frac{\Gamma(\alpha)}{4 \pi^{\alpha +1}\Gamma(2\alpha)}  \frac{\partial}{\partial \mu^2} \frac{\partial}{\partial s} \sum_\pm  \sum_{J=0}^\infty (2J+2\alpha)\frac{\Gamma(J+2\alpha)}{\Gamma(J+1)}(J-h_\pm)^s|_{s=0}~.
\end{equation}
We will begin by regulating the expression $\sum_{J=0}^\infty (2J+2\alpha)\frac{\Gamma(J+2\alpha)}{\Gamma(J+1)}(J-a)^s$, where $a$ is a placeholder for $h_\pm$.
We expand the summand in powers of $J$ and isolate the term that produces a logarithmic divergence, which we will fix using dimensional regularization.
Equivalently, this procedure amounts to expanding the sum in terms of zeta functions, in which case the only singularity arises from the pole of $\zeta(1)$
\be
\sum_{J=0}^\infty (2J+2\alpha)\frac{\Gamma(J+2\alpha)}{\Gamma(J+1)}(J-a)^s = \alpha_0 \sum_J J^{d-1+s} + \cdots + \beta \sum_J J^{d-5+s}+ \cdots~,
\ee
where $\b$ marks the coefficient multiplying $\z(1)$, and all the coefficients depend on $a,d$, and $s$. We will make use of the fact that 
\be
\frac{\partial}{\partial s}\beta(a,d,s)\sum_J^\infty J^{-1-\epsilon+s}\bigg|_{s=0} = \frac{\beta(a,d,0)}{\epsilon^2} + \frac{1}{\epsilon}\frac{\partial \beta(a,d,0) }{\partial s} + \mathcal{O}(1)~,
\ee
is the only $1/\epsilon$ pole and we can ignore all other terms. So it remains for us to find $\beta(a,d,s)$, which we can easily do since we have the explicit expression for everything involved. We can then expand around $\epsilon = 0$ and keep the first two terms 
\begin{align}
    \beta(a,d,s)&=\frac{1}{2880} \big(-240 d^7 + 15 d^8 + d^6 (1670 - 120 a s) + 24 d^5 (-277 + 50 a s) \nonumber \\
    & -120 d^3 (217 - 96 a s + 16 a^2 (-1 + s) s) + 5 d^4 (3307 - 1008 a s + 72 a^2 (-1 + s) s) \nonumber \\
    & + 48 d (-274 + 215 a s - 80 a^2 (-1 + s) s + 20 a^3 s (2 - 3 s + s^2)) \nonumber \\
    & - 20 d^2 (-1249 + 750 a s - 198 a^2 (-1 + s) s + 24 a^3 s (2 - 3 s + s^2)) \nonumber \\
    & + 240 (12 - 12 a s + 6 a^2 (-1 + s) s - 2 a^3 s (2 - 3 s + s^2) + a^4 s (-6 + 11 s - 6 s^2 + s^3)) \big)~.
\end{align}
By inspection, we can see that every factor of $a$ in the above expression appears multiplied by $s$. Thus, when we set $s = 0$, all dependence on $a$ drops out, and in particular no terms proportional to $\mu^{2}$ survive. Differentiating with respect to $\mu^{2}$ therefore annihilates these contributions. Consequently, we only need to retain the $\mathcal{O}(1)$ part of $\beta'$. This can be done explicitly, after which we simply replace $a$ with $h_{\pm}$ and differentiate with respect to $\mu^{2}$. This finally gives us in $d=4$
\be
\frac{\partial}{\partial \mu^2}\sum_\pm  \frac{1}{\epsilon}\frac{\partial \beta(h_\pm,4,0) }{\partial s}   = \frac{2-\mu^2}{\epsilon}~.
\ee
Reinstating an overall normalization factor, we finally extract the $1/\epsilon$ to be
\be
\frac{2-\mu^2}{8 \pi^{2} }\frac{1}{\epsilon}~.
\label{eq:divofg}
\ee

\subsection{Divergence of $G'(h)|_{h = 1}$}

To extract the divergence of the derivative of the propagator $G_E'(\sigma)$, we now need to differentiate with respect to $\sigma$, which is straightforward given that we have the relation
\be
\frac{\partial}{\partial\s}C^\alpha_J(\sigma) = 2\alpha C^{\alpha+1}_{J-1}(\sigma)~.
\ee
Using the above, we obtain
\begin{align}
G_E'(\sigma)|_{\sigma=1} &=  \frac{\Gamma(\alpha)}{2 \pi^{\alpha +1}}\sum_{J=0}^\infty\frac{J+\alpha}{J(J+2\alpha)+\mu^2}  2\alpha C^{\alpha+1}_{J-1}(\sigma = 1)~ \nonumber \\
&= \frac{\Gamma(\alpha)}{2 \pi^{\alpha +1}}\sum_{J=0}^\infty\frac{2\alpha(J+\alpha)}{J(J+2\alpha)+\mu^2}   \frac{\Gamma(J+2\alpha+1)}{\Gamma(J)\Gamma(2(\alpha +1))}~.
\end{align}
We can now repeat the same procedure as in the previous appendix \ref{app:divofprop} to extract the $1/\epsilon$ divergence of $G'(\sigma)$. After all is said and done, we obtain
\be
\frac{\mu^2(\mu^2-2)}{32 \pi ^2} \frac{1}{\epsilon}~.
\label{eq:divofGprime}
\ee
Combining \eqref{eq:divofg} and \eqref{eq:divofGprime} into the relation \eqref{eq:divofstresstensor}, we can now express the divergence of the stress tensor as 
\be
 -\frac{1}{2} U \cdot V ((\mu^2-6\xi) G_\epsilon(1) +2G'_\epsilon(1)) = U\cdot V \frac{m^2(m^2 + 12\xi-2)}{32\pi^2}  \frac{1}{\epsilon}~.
\ee
For both the conformal case $m^2 + 12\xi = \mu^2 = 2$, and for the massless case $m^2 =0$, the $1/\epsilon$ contribution vanishes and there are no divergences.

We note, of course, that in the conformal case the one-point function of the stress tensor vanishes identically, since by symmetry it is proportional to the metric, and is therefore proportional to its trace, which vanishes in a conformal theory. However, even when the stress tensor is inserted alongside other operators and is no longer proportional to the metric, the divergence still vanishes.


\bibliographystyle{JHEP}
\bibliography{refs}


\end{document}